\documentclass[prx, twocolumn, longbibliography, superscriptaddress, urlcolor=blue, linkcolor=blue]{revtex4}
\usepackage{amsmath,amsfonts, amsthm}
\usepackage{graphicx, subfigure}
\usepackage{subfigure}
\usepackage{comment,color}

\newcommand{\cblue}[1]{\textcolor{black}{#1}}

\newcommand{\s}{\text{s}}
\newcommand{\GSD}{\text{GSD}}

\begin{document}

\def\c#1{\mathbb{#1}}
\def\t#1{\widetilde{#1}}
\def\braket#1{\left(#1\right)}
\def\sb#1{\left[#1\right]}
\def\bra#1{\left\langle#1\right|}
\def\ket#1{\left|#1\right\rangle}
\def\tx#1{{\rm{#1}}}
\def\pa{\partial}
\def\ml#1{${#1}$}
\def\ma#1{\begin{align}
#1
\end{align}}

%\renewcommand{\section}[1]{ {\it #1} }
%\title{Mechanisms and bulk dynamical effective actions for Abelian symmetry-protected topological phases:  beyond Chern-Simons and BF theories}
%\title{Multi-kink Berry phase  and vortex condensation mechanism for symmetry-protected topological states: Bulk-edge field theory beyond Chern-Simons and BF theories}
%\title{Binding charge to domain-wall intersection point and\\ domain-wall
%condensation induced symmetry-protected topological states}
\title{Multi-kink topological terms and
charge-binding domain-wall condensation induced symmetry-protected topological states: beyond-Chern-Simons/BF field theories}
%Field theories with anomalous symmetry}
%and beyond-Chern-Simons/BF field theories
%%%%%%%%%%%%%%%%%%%%%%%%
%\title{Disordering superfluids into symmetry-protected topological states: \\
%Multi-kink Berry phase topological term and vortex condensation mechanism}

\author{Zheng-Cheng Gu} \email{zgu@perimeterinstitute.ca}
\affiliation{Perimeter Institute for Theoretical Physics, Waterloo, ON, N2L 2Y5, Canada}

\author{Juven C. Wang}  \email{juven@mit.edu}
\affiliation{Department of Physics, Massachusetts Institute of Technology, Cambridge, MA 02139, USA}
\affiliation{Perimeter Institute for Theoretical Physics, Waterloo, ON, N2L 2Y5, Canada}

\author{Xiao-Gang Wen} \email{wen@dao.mit.edu}
\affiliation{Department of Physics, Massachusetts Institute of Technology, Cambridge, MA 02139, USA}
\affiliation{Perimeter Institute for Theoretical Physics, Waterloo, ON, N2L 2Y5, Canada}

\begin{abstract}
Quantum-disordering a discrete-symmetry breaking state by condensing
domain-walls can lead to a trivial symmetric insulator state.  In this work, we
show that if we bind a 1D representation of the symmetry (such as a charge) to
the intersection point of several domain walls, condensing such modified
domain-walls can lead to a non-trivial symmetry-protected topological (SPT)
state.  This result is obtained by showing that the modified domain-wall
condensed state has a non-trivial SPT invariant -- the symmetry-twist dependent
partition function.  We propose two different kinds of field theories that can
describe the above mentioned SPT states. The first one is a
Ginzburg-Landau-type non-linear sigma model theory, but with an additional
multi-kink domain-wall topological term.  Such theory has an anomalous $U^k(1)$
symmetry but an anomaly-free $Z_N^k$ symmetry.  The second one is a gauge
theory, which is beyond Abelian Chern-Simons/BF gauge theories. We argue that
the two field theories are equivalent at low energies. After coupling to the
symmetry twists, both theories produce the desired SPT invariant.

\end{abstract}

%\date{\today}
\maketitle

\tableofcontents

\section{Introduction}

\subsection{SPT states and their effective field theories}

Recently, it has been realized that many-body ground states can be divided into
two classes:\cite{Xietop} long-range entangled (LRE) states and short-range
entangled (SRE) states. The LRE states can belong to many different phases that
correspond to topologically ordered phases.\cite{Wentop,Wenrig} When there is a
global symmetry (described by a group $G$), even SRE states can belong to many
different phases, and these phases are called symmetry-protected topological
(SPT) states.\cite{GuSPT,PollmannSPT1,PollmannSPT2,XieSPT1,XieSPT2,NorbertSPT}
A large class of bosonic SPT states whose boundary has a pure ``gauge
anomaly''\cite{duality,dualityfermion,WenSPT1} can be systematically classified
via group cohomology classes $H^{d+1}(G,\R/\Z)$.\cite{XieSPT3,XieSPT4,XieSPT5}
All these SPT states can be realized by exactly-soluble lattice non-linear
$\sigma$-model with the symmetry group $G$ as the target space plus a $2\pi$
quantized topological $\theta$-term.  They can also be realized by
exactly-soluble lattice Hamiltonians that contain seven-body interactions.  In
addition, bosonic SPT states whose boundary has a ``gauge gravitational mixed
anomaly'' can all be realized by lattice non-linear $\sigma$-model with
$SO_\infty\times G$ as the target space and with a $2\pi$ quantized topological
$\theta$-term.\cite{WenSPT2} The potentially possible SPT invariants of the
first and the second classes of SPT states can also be studied directly via
cobordism theory,\cite{KSPT1,KSPT2,KSPT3,JuvenSPT1} but the cobordism theory
does not give rise to a realization of the SPT states.

Many of the SPT states protected by discrete group symmetry can also be realized by condensing domain walls in
symmetry breaking states, if we decorate the domain walls with lower dimensional
SPT states and/or invertible topologically ordered
states.\cite{SenthilBF,CLV1407,WenSPT3,WenSPT2}
In this work, we will realize some additional SPT states by condensing domain
walls, such that the intersection point of
several domain walls carries the quantum number of the unbroken symmetries.  More
general SPT states protected by discrete group symmetry can be obtained by decorating the intersection lines (or
surfaces) of several domain walls with 1D (or 2D) SPT states (as indicated by the
Kunneth formula for the group cohomology \cite{WenSPT3,WenSPT2}).

In addition to the above systematic constructions of all the bosonic SPT
phases, people have also developed many field theory realizations for some
special simple SPT states (under the name of bosonic topological insulator
(BTI)\cite{SenthilBF,MaxBTI,PengBTI,WangBTI,XuBTI1,XuBTI2,XuBTI3,LiuBTI,LesikBTI,Ye:2014oua})
which lead to some simple physical pictures and mechanisms for bosonic SPT
states.  Due to the incompressibility of topological phases, it is sufficient
to only consider quantum fluctuations of collective modes at low energies and
long wave-lengths, e.g., density and current fluctuations. Such an approach is
the so-called ``hydrodynamical approach'' or effective quantum field theory for
topological phases.  The field theory realizations of SPT states belong to this
approach.

Historically, the ``hydrodynamical approach'' turns out to be extremely
powerful to understand the underlying physics of topological phases. For
example, the fractional quantum Hall effect (FQHE) can be understood by the
Ginzburg-Landau  Chern-Simons theory\cite{CS0} or more systematically by pure
Chern-Simons theory\cite{CS1,CS2,CS3,CS4,CS5,CS6,CS7}.  Those bulk dynamic
effective theories that capture the low energies and long wave-length physics
are also very useful to study phase transitions among different topological
phases, e.g, phase transitions between FQHE at different filling fractions.
Thus, the bulk dynamical Chern-Simons action approach to FQHE phases can be
viewed as the Ginzburg-Landau action approach to symmetry breaking phases.
Therefore, it is very natural to ask what is the ``hydrodynamical approach''
to SPT states.

Very recently, Chern Simons/BF theories have been proposed
\cite{SPTCS1,SPTCS2,SPTCS3,MooreBF,SenthilBF,{Ye:2013upa},{Wang:2014tia}} as
bulk dynamical effective actions to describe 2D/3D bosonic SPT states protected
by Abelian symmetry group (the so-called Abelian SPT states).  Nevertheless, it
has been pointed out\cite{SPTCS3} that the \cblue{Abelian} Chern Simons/BF
theory approach is incomplete. Thererfore, a much more general theoretical
framework for bulk dynamical actions of SPT states is very desired.  In this
paper, we will focus on the mechanisms and bulk dynamical effective actions for
bosonic SPT states \cblue{with finite Abelian group symmetry} within group
cohomology classification. We propose a class of new topological actions to
characterize bosonic Abelian SPT states in arbitrary dimensions that are beyond
\cblue{Abelian} Chern-Simons/BF theory. We will show that such a class of
generalized topological actions serves as a complete description for bosonic
Abelian SPT states in 1D and 2D.  In 3D, there are still some Abelian SPT
states beyond the proposed bulk dynamical effective action; however, we believe
that the basic principle and method developed in this paper are still
applicable. We will leave these studies for future work. It is also worthwhile
to mention that in a parallel work\cite{Ye:2014oua}, a bulk dynamical effective
action for Abelian SPT states beyond group cohomology classifications is also
proposed. In principle, the ``hydrodynamical approach'' can also be generalized
into interacting fermionic systems.
% and even with non-Abelian symmetry groups.

\subsection{Summary of results}

\begin{figure}[t] %!htbp
{\includegraphics[width=.48\textwidth]{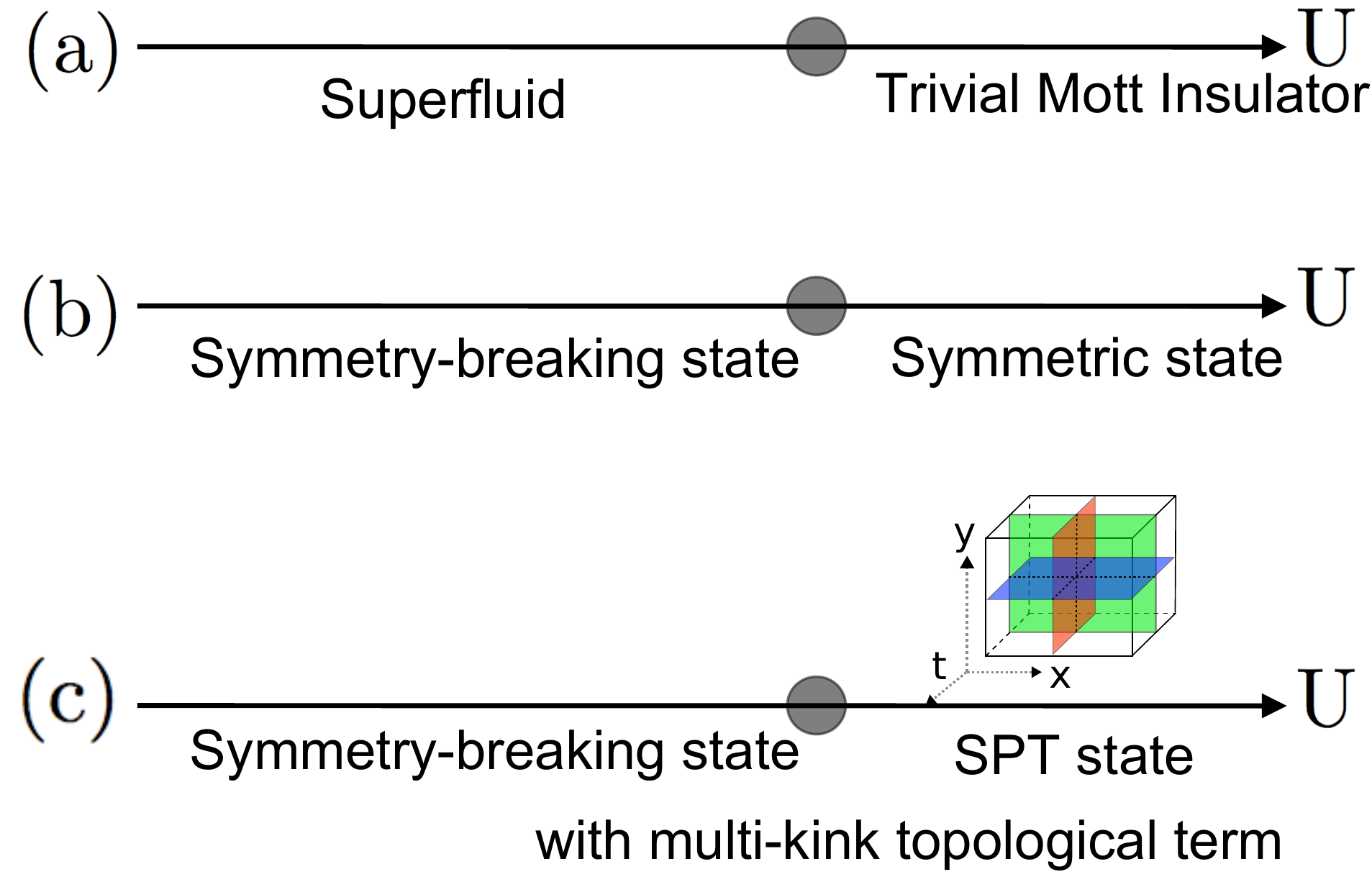}
\caption{
%\textbf{Change "superfliud" to "symmetry breaking", "Mott insulator" to "symmetric phase", "Barry phase" to "topological term" in the figure}
(a) Disordering a $U(1)$-symmetry breaking superfluid with an action
by condensing the vortices, e.g., tuning some coupling constant U to increase the charge repulsion. \cite{{Fisher-Lee},{Dasgupta-Halperin},{Nelson}}.
(b) Disordering a discrete-symmetry breaking state by condensing the domain
walls. The gray region qualitatively indicates the phase transition region,
such as a critical point or a different phase.
(c) In this work, we generalize the previous process by condensing domain walls
with a multi-kink topological terms.  We obtain nontrivial SPT states with SPT invariants listed in
Table \ref{table:field_theory}.
%The SPT state can be detected by the
%symmetry-twist multi-kink intersecting configuration predicted in
%Ref.\cite{JuvenSPT1}.
}
\label{Fig_multi_kink_SF_SPT}}
\end{figure}

%\begin{widetext}
%\onecolumngrid
\begin{table*}[t]
 \begin{tabular}[t]{|c||c|c|c|}
 \hline
  & \begin{minipage}[t]{2.2in}
%SPT intrinsic bulk field theory with multi-kink topological term (approached from the symmetry-breaking state):\\
Ginzburg-Landau NL$\sigma$M\end{minipage}& \begin{minipage}[t]{1.5in}
%SPT intrinsic bulk field theory with gauge fields (approach to the insulator):\\
Dynamical gauge theory\end{minipage}
 &\begin{minipage}[t]{1.5in} SPT invariants:\\ Probed field theory\\ %or dynamically gauged field theories
 \end{minipage}\\
 \hline\hline
1D & \begin{minipage}[t]{2.2in} ${\frac{ \chi}{2}} (\partial_\mu \theta^I)^2+\frac{\ii}{2}C_{IJ}\varepsilon^{\mu\nu}
\partial_\mu \theta^I \partial_\nu \theta^J$ \end{minipage} &
\begin{minipage}[t]{1.4in}
 $\frac{\ii}{2\pi}\varepsilon^{\mu\nu}\cblue{b_\mu^I  \partial_\nu a^I}+$\\
$\cblue{\frac{-\ii}{2} C_{IJ}}\varepsilon^{\mu\nu}b_\mu^I b_\nu^J  $  %\cblue{+\dots}
\end{minipage} &
\begin{minipage}[t]{1.2in}
%$\frac{\ii \cblue{N_I}}{2\pi}\varepsilon^{\mu\nu} B^I \partial_\nu A^I_\mu  +$ \\
$ \cblue{\frac{-\ii}{2} C_{IJ}}\varepsilon^{\mu\nu} A^1_\mu A^2_\nu$ % \cblue{+\dots}
\end{minipage}\\
\hline
2D  & \begin{minipage}[t]{2.2in} ${\frac{\chi}{ 2}}(\partial_\mu \theta^I)^2$ +
$\frac{\ii }{3} C_{IJK}\varepsilon^{\mu\nu\lambda}\partial_\mu \theta^I \partial_\nu \theta^J
\partial_\lambda \theta^K $ \end{minipage}  &
\begin{minipage}[t]{1.5in}
 $\frac{\ii \varepsilon^{\mu\nu\lambda}}{2\pi} b_\mu^I \partial_\nu a_\lambda^I +$ \\
$\frac{\ii C_{IJK}}{3} \varepsilon^{\mu\nu\lambda} b_\mu^I b_\nu^J \lambda_b^K$  % \cblue{+\dots}
\end{minipage}
&
\begin{minipage}[t]{1.5in}
%$\frac{\ii N_I}{2\pi}\varepsilon^{\mu\nu\lambda}B_\mu^I\partial_\mu A_\nu^I+$ \\
$\frac{\ii C_{IJK}}{3} \varepsilon^{\mu\nu\lambda} A_\mu^1 A_\nu^2 A_\lambda^3$ %\cblue{+\dots}
\end{minipage}
\\
\hline
3D  &
\begin{minipage}[t]{2.7in} ${\frac{\chi}{ 2}}(\partial_\mu \theta^I)^2$ +
$\frac{\ii }{4} C_{IJKL}\varepsilon^{\mu\nu\lambda \sigma}\partial_\mu \theta^I\partial_\nu \theta^J \partial_\lambda \theta^K \partial_\sigma \theta^L$ \end{minipage}
&
\begin{minipage}[t]{1.5in}
  $\frac{\ii \varepsilon^{\mu\nu\rho\sigma}}{4\pi} b_\mu^I \partial_\nu a_{\sigma\rho}^I +$\\
$ \frac{-\ii C_{IJKL}}{4} \varepsilon^{\mu\nu\sigma\rho} b_\mu^I b_\nu^J b_\sigma^K b_\rho^L $ %\cblue{+\dots}
\end{minipage}
&
\begin{minipage}[t]{1.5in}
%$\frac{\ii N_I}{4\pi}\varepsilon^{\mu\nu\rho\sigma}B_{\mu\nu}^I\partial_\rho A_\sigma^I +$ \\
$\frac{-\ii C_{IJKL}}{4} \varepsilon^{\mu\nu\rho\sigma} A_\mu^I A_\nu^J A_\rho^K A_\sigma^L$ %\cblue{+\dots}
\end{minipage}\\
\hline
\hline
 \end{tabular}
 \caption{
First column: the $U^k(1)$ non-linear $\sigma$-model (NL$\sigma$M) realization of the $
Z_{N_1}\times Z_{N_2}\times Z_{N_3}\times \cdots$ SPT states
in the $\chi<\chi_c$ disordered limit.
%(in the strong coupling $g\to \infty$ limit).
The additional multi-kink topological term
(bi-kink for 1+1D, tri-kink for 2+1D, quad-kink for 3+1D, etc) are listed.  The
phase fluctuating term $\partial_\mu\theta^I \equiv\partial_\mu \theta^I_{\s} + b_\mu^I$ contain a smooth piece
$\partial_\mu \theta^I_{\s}$ and a singular piece $b_\mu^I$.  Here $C_{IJ \dots}$ is
a totally anti-symmetric tensor, with: $C_{12}=\frac{1}{(2 \pi)} \frac{N_1 N_2\;
p_{\text{II} }}{N_{12}}$, $C_{123}=\frac{1}{(2 \pi)^2 2!} \frac{N_1 N_2 N_3\;
p_{\text{III} }}{N_{123}}$,  $C_{1234}=\frac{1}{(2 \pi)^3 3!}  \frac{N_1 N_2
N_3 N_4\; p_{\text{IV} }}{N_{1234}}$, etc., with ${N_{12 \dots}} \equiv \gcd(N_1,N_2, \dots)$.
Second column: the dynamical gauge theory
realization of   the $ Z_{N_1}\times Z_{N_2}\times Z_{N_3}\times \cdots$ SPT
states.  The important global constraints on the fields are not specified,  % in our Lagrangian
moreover we need to well-define the SPT path integral more than just the SPT Lagrangian;
we will discuss this issues of path integral in depth in Sec.\ref{sec:partition-GSD}.
% but those $\dots$ terms can be important to fully characterize the bulk SPT
% state.
Third column: the SPT invariants after integrating out the matter fields. Here
the non-dynamical flat $A^I$ field describes the $Z_{N_I}$-symmetry twist, which
satisfies $\oint A^I_\mu \dd x^\mu = 0 $ mod $2\pi/N_I$.  The main result of our
work is that the field theories in the first and the second columns are
equivalent at low energies at the $\chi<\chi_c$ disordered limit.  We can derive their SPT invariants by integrating
out the matter field. The SPT invariant is of the form: $\int {\dd^d x}
\frac{\ii C_{I_1 I_2 \dots I_{d}}}{d} \varepsilon^{\mu\nu \dots \sigma}
A_\mu^{I_1} A_\nu^{I_2} \dots A_\sigma^{I_{d}}$ given in \cite{JuvenSPT1}.
 }\label{table:field_theory}
 \end{table*}
%\end{center}
%\end{widetext}

\subsubsection{A mechanism of SPT states}

Let us start by summarizing the mechanism that generate SPT states at intuitive
level.  It is well-known that if we disorder a discrete-symmetry breaking state
by condensing domain walls, we can obtain a symmetry restored state.
%, see the pioneer work\cite{{Fisher-Lee},{Dasgupta-Halperin},{Nelson}} and
%Ref.\cite{Zee:2003mt} for a field theory approach.
Our approach is basically analogous to this line of thinking, except that we
generalize the approach by including additional multi-kink topological terms to
the domain walls, see Fig.\ref{Fig_multi_kink_SF_SPT}.

\cblue{There are two ways to view the multi-kink topological terms: the space
picture and the space-time picture.  In the space picture, we create the
symmetry-breaking domain walls and trap some charges (not fractionalized) of the
remained unbroken symmetry at the intersecting points, then we proliferate and
condense the domain walls to restore the broken symmetry.
%The vortices can be viewed as the singular core of U(1) quantum phase.  The
%vortex theory description is the the dual description of the U(1) quantum
%phase theory.
On the other hand, in the spacetime picture, we have an intersecting profile that
contributes a nontrivial phase to the path integral (see
Fig.\ref{Fig_multi_kink_SF_SPT} (b)), and we then disorder the symmetry
breaking state with such nontrivial multi-kink topological term.  As we will
show explicitly and quantitatively using field theories, both processes lead
to a nontrivial SPT state.}

Using the above domain-wall condensation picture, we also obtain two kinds of
field theory realization of the corresponding $ Z_{N_1}\times Z_{N_2}\times
Z_{N_3}\times \cdots$ SPT states (see Table \ref{table:field_theory}).  The
first one is a $U^k(1)$ non-linear $\sigma$-model with a multi-kink topological
term.  The second one is a dynamical gauge theory that is beyond Abelian
Chern-Simons/BF theory. \cblue{Throughout the whole paper, we will implement
the Euclidean spacetime approach with the Euclidean time $t_E = \ii t$ as the
Minkowski time Wick-rotated by an imaginary $\ii$. We define the derivative
$\partial_0$ as $\partial_{t_E}$. We choose the Euclidean spacetime for the
future convenience of the lattice regularization.}

In the first column of Table \ref{table:field_theory}, we list the $U^k(1)$
non-linear $\sigma$-models with the multi-kink topological terms of the form
$\frac{\ii}{d}C_{IJK \dots}\varepsilon^{\mu\nu\lambda \dots}\partial_\mu \theta^I\partial_\nu \theta^J \partial_\lambda \theta^K \dots$ with $C_{IJK \dots}$ a fully anti-symmetric tensor and $d$ the spacetime dimension.  In the
second column of Table \ref{table:field_theory}, we list the gauge theory
realization of the same $Z_{N_1}\times Z_{N_2}\times Z_{N_3}\times \cdots$ SPT
states.  Our local field theories in the first and the second columns can
produce the desired SPT invariants dictated by group cohomology
\cite{JuvenSPT1} (after integrating out the dynamical fields).  We list the SPT
invariants in the third column of Table \ref{table:field_theory}.

\subsubsection{Field theory with anomalous $U(1)$ symmetry}

We stress that although the proposed $U^k(1)$ non-linear $\sigma$-model with
the multi-kink topological terms formally has a $U^k(1)$ global symmetry
$\th_I(x^\mu) \to \th_I(x^\mu) +\Delta f_I$.
Due to the presence of multi-kink topological terms, the $U^k(1)$
global symmetry is actually anomalous, \ie cannot be realized by an
on-site-symmetry\cite{WenSPT1} in any lattice regularization of the field
theories.  Or more precisely, the $U^k(1)$ non-linear $\sigma$-models have
anomalous $U^k(1)$ symmetry if the multi-kink topological terms are quantized
as  $C_{12}=0 \text{ mod } \frac{1}{(2 \pi)}$ in 1+1D, $C_{123}=0 \text{ mod }
\frac{1}{(2 \pi)^2 2!} $ in 2+1D, and  $C_{1234}=0 \text{ mod } \frac{1}{(2
\pi)^3 3!} $ in 3+1D.

Here we use a $1+1$D example to explain the above statement (the higher
dimensional cases can be understood in a similar way). Let us consider an ideal
experiment by inserting a $2\pi$ flux corresponding to the first $U(1)$
symmetry through a closed 1D ring, the bi-kink topological term
$\frac{\ii}{2}C_{IJ}\varepsilon^{\mu\nu} \partial_\mu \theta^I \partial_\nu
\theta^J$ will induce a charge $2\pi C_{12}$ associate with the second $U(1)$
symmetry.  So if the $2\pi C_{IJ}$ is not an integer, the $U^k(1)$ non-linear
$\sigma$-model does not even have the $U^k(1)$ symmetry at quantum level.  When
$2\pi C_{IJ} \in \Z$, the  $U^2(1)$ symmetry is anomalous, since adding the
flux of the first $U(1)$ can cause a non-conservation of the second $U(1)$.

The above charge pumping phenomena via flux insertion can happen on a boundary
of a $2+1$D system, where an integer charge is created in the bulk and the
total $U^k(1)$ charges are conserved.

However, the above charge pumping phenomena cannot happen in a strict $1+1$D
system with on-site $U^k(1)$ symmetry.  This is because the on-site $U^k(1)$
symmetry is gaugable (\ie we can add $U(1)$-flux without breaking the $U^k(1)$
symmetry).  The presence of the charge pumping phenomena implies that, at
quantum level, the $U^k(1)$ symmetry is broken by the $U(1)$-flux, which in
turn implies that the $U^k(1)$ symmetry is anomalous (or non-on-site).  Or in
other words, in a strict $1+1$D system with $U(1)^2$ on-site symmetry, $C_{12}$
must vanish.

On the other hand, if the $2\pi C_{12} = 0 \text{ mod } \frac{N_1
N_2}{N_{12}}$, the $Z_{N_1}\times Z_{N_2}$ subgroup of the $U(1)^2$
correspond to an anomaly-free symmetry (\ie an on-site symmetry).  This is
because the $2\pi$-flux of $U(1)$ induce a charge $2\pi C_{12} = \frac{N_1
N_2}{N_{12}}\times $ integer, which is essentially trivial since $Z_{N_2}$
charge is only conserved mod $N_2$. Therefore, the $Z_{N_1}\times Z_{N_2}$
subgroup of the $U(1)^2$ is no anomalous. The $U^2(1)$ non-linear
$\sigma$-model describe a system with $Z_{N_1}\times Z_{N_2}$ on-site symmetry,
if $2\pi C_{12} = 0 \text{ mod } \frac{N_1 N_2}{N_{12}}$.

Similarly, the $U^k(1)$ non-linear $\sigma$-model have a $Z_{N_1}\times
Z_{N_2}\times Z_{N_3}\times \cdots$ on-site symmetry only when proper quantized
values is assigned for $C_{IJK \dots}$. For example, in $1+1$D, $2+1$D and
$3+1$D, we require that $C_{12}=\frac{1}{(2 \pi)} \frac{N_1 N_2\; p_{\text{II}
}}{N_{12}}$, $C_{123}=\frac{1}{(2 \pi)^2 2!} \frac{N_1 N_2 N_3\; p_{\text{III}
}}{N_{123}}$ and  $C_{1234}=\frac{1}{(2 \pi)^3 3!} \frac{N_1 N_2 N_3 N_4\;
p_{\text{IV} }}{N_{1234}}$, where $ p_\text{I}, p_\text{II}, p_\text{III} \in
\Z$.

\subsection{Organization of the paper}

The rest of the paper is organized as follows: In Section II, we briefly review
how to use SPT invariants to define SPT states. In Section III, we propose a
bulk dynamical effective action to describe 1+1D bosonic Abelian SPT states and
use it to derive the corresponding SPT invariants. In Section III, we briefly
review the Chern-Simons action approach for 2+1D bosonic Abelian SPT states and
discuss its limitation. In Section IV, we compute the SPT invariants for 2+1D
$Z_{N_1}\times Z_{N_2}\times Z_{N_3}$ SPT state and propose a bulk dynamical
effective action to describe such 2+1D SPT states. In Section VI, we generalize
our results to 3+1D bosonic Abelian SPT states and propose a bulk dynamical
action beyond BF theory. In Section VII, edge theories for Abelian SPT states
beyond Chern-Simons/BF actions are discussed via a standard dimension reduction
scheme. Finally, there are conclusion remarks and discussions for future
directions.

\cblue{
In Appendix \ref{sec:disordersf},
we review the derivation of disordering the superfluid state to the Mott insulator, see the
pioneer work\cite{{Fisher-Lee},{Dasgupta-Halperin},{Nelson}} and Ref.\cite{Zee:2003mt,Wenbook}.
%In this work, we generalize the above procedure using field theory approach.
%We disorder the symmetry-breaking state with additional nontrivial multi-kink topological terms,
%and, by condensing the quantum vortices, we can generate a symmetry-protected insulator: the SPT state.
%The main result of our field theory is summarized in Table \ref{table:field_theory}.
In Appendix \ref{derivation}, we provide an explicit calculation of an effective bulk action of SPT state.
In Appendix \ref{sec:partition-GSD}, we verify that the partition function with the proposed SPT action has the GSD=1.
In Appendix \ref{non-semi-Simple}, we provide some words of caution by comparing our effective action of SPT state
to topological gauge theories with non-semi-simple Lie algebra.
In Appendix \ref{GSD}, we compute the edge mode GSD by counting the degenerate zero modes.}

%\cblue{{\bf Juven: comment about the Euclidean/Minkowski? $\partial_0$ for the Euclidean time?}}

\section{A review of SPT states defined by SPT invariants}

It has been shown that SPT states (within group cohomology or beyond group cohomology classifications) can be
probed or even defined through the so-called SPT invariants\cite{WenSPT3,JanetSPT1} that may
completely characterize different SPT states.  In this section, we
will review and discuss such a point of view.

\subsection{Universal wavefunction overlap: a complete SPT
invariant for SPT orders}

%To begin,
We start from reviewing the results of the SPT invariants in
\Ref{JanetSPT1}, using 2+1D systems as examples.  It was conjectured that the
degenerate ground states $|\Psi_\al\>$, $\al=1,2,\cdots$, of a 2+1D topological phase
on a torus have the following properties:\cite{Kongtop}
\begin{align}
S_{\al\bt}\ e^{-f_S L^2+o(L^{-1})} &= \<\Psi_\al|\hat S |\Psi_\bt\>
\nonumber\\
T_{\al\bt}\ e^{-f_T L^2+o(L^{-1})} &= \<\Psi_\al|\hat T |\Psi_\bt\>
\end{align}
where $\hat S$ is the $90^\circ$ rotation operation $(x,y)\to(-y,x)$ and
$\hat T$ is the Dehn twist rotation operation $(x,y)\to(x+y,y)$.  It was
conjectured that while the complex numbers $f_S$ and $f_T$ are not universal,
the complex matricies $S_{\al\bt}$ and $T_{\al\bt}$ are universal.  $S_{\al\bt}$
and $T_{\al\bt}$ can change only via phase transitions.  Thus we can use them
to characterize different topological orders.  In fact, we believe that
$S_{\al\bt}$ and $T_{\al\bt}$ completely define 2+1D topological ordered phases with
gappable edges.

Can we use the similar idea to completely define 2+1D SPT order?
The wavefunction overlap for SPT state also has the following
universal structure
\begin{align}
S\ e^{-f_S L^2+o(L^{-1})} &= \<\Psi_0|\hat S |\Psi_0\>
\nonumber\\
T\ e^{-f_T L^2+o(L^{-1})} &= \<\Psi_0|\hat T |\Psi_0\>
\end{align}
where the $1\times 1$ unitary matrices $S$ and $T$ are universal.  In fact
\ml{S=T=1}, due to the trivial bulk excitations in SPT state.  Thus  $S$ and
$T$ are trivial and could not be used to distinguish different SPT states.

\begin{figure}[tb]
\centering
\subfigure[\label{ltrans}]{
\includegraphics[width=.22\textwidth]{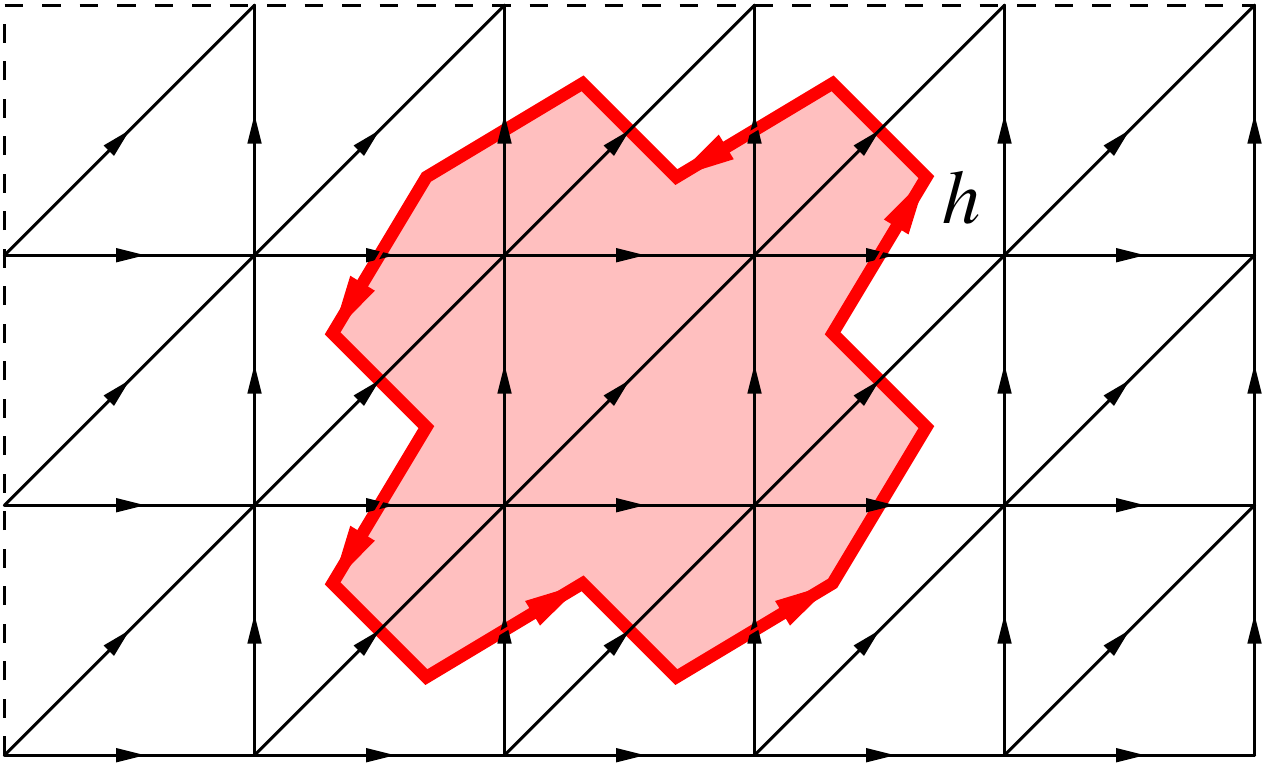}}~
\subfigure[\label{hxhy} ]{
\includegraphics[width=.22\textwidth]{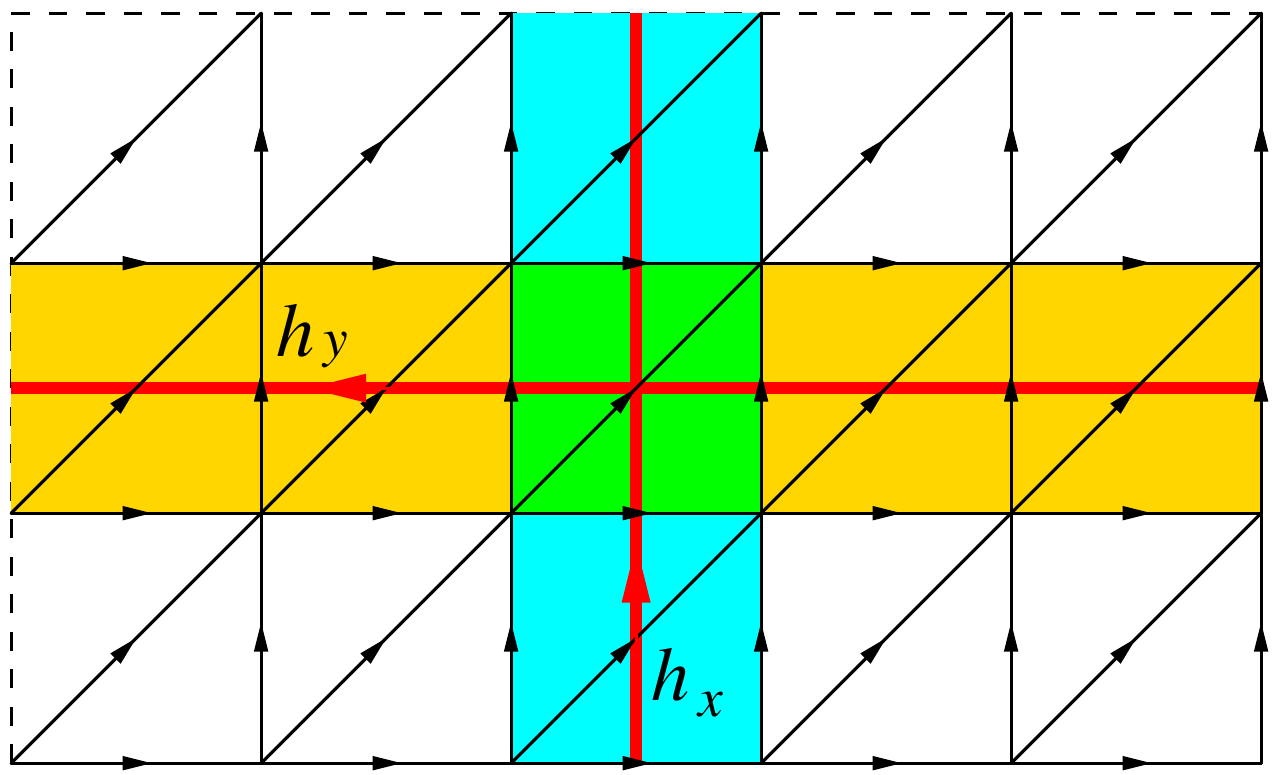}}
\caption{
(a) Symmetry twist along the boundary $\prt R$
is generated by the symmetry transformation that act only within $R$.
(b) The symmetry twist $h_x,h_y$ on torus gives rise to the
twisted ground state  \ml{|\Psi_{(h_x,h_y)}\>}.
}
\end{figure}

To obtain a non-trivial
wavefunction overlap, we introduce symmetry twist:
a symmetric transformation generated by $h\in G$ within the region $R$.
The Hamiltonian is not invariant under such a local symmetry transformation
(see Fig. \ref{ltrans}):
\ma{H=\sum H_{ijk} \to
H_h=\sum_\text{in $R,\bar{R}$} H_{ijk} +\sum_\text{on $\prt R$} H^{h}_{ijk}
}
where $H_{ijk}$ acts on sites $i,j,k$ and $ H^{h}_{ijk}$ is on the boundary
of $R$, $\prt R$, if the sites $i,j,k$ are not all on one side of $\prt R$.  We
call \ml{\sum_\text{on $\prt R$} H^{h}_{ijk} } the \ml{h}-symmetry twist.

\begin{figure}[tb]
\centering
\includegraphics[height=1in]{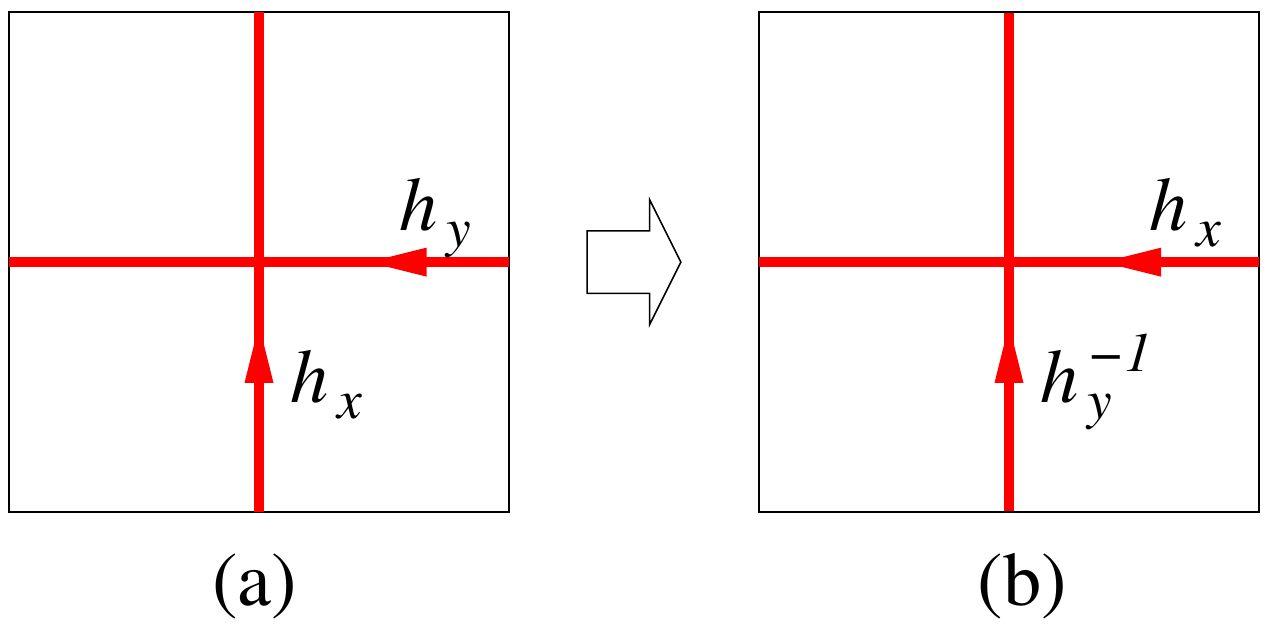}
\caption{ $\hat S$-move is $90^\circ$ rotation.  }
\label{Smov}
\end{figure}

\begin{figure}[tb]
\centering
\includegraphics[height=1in]{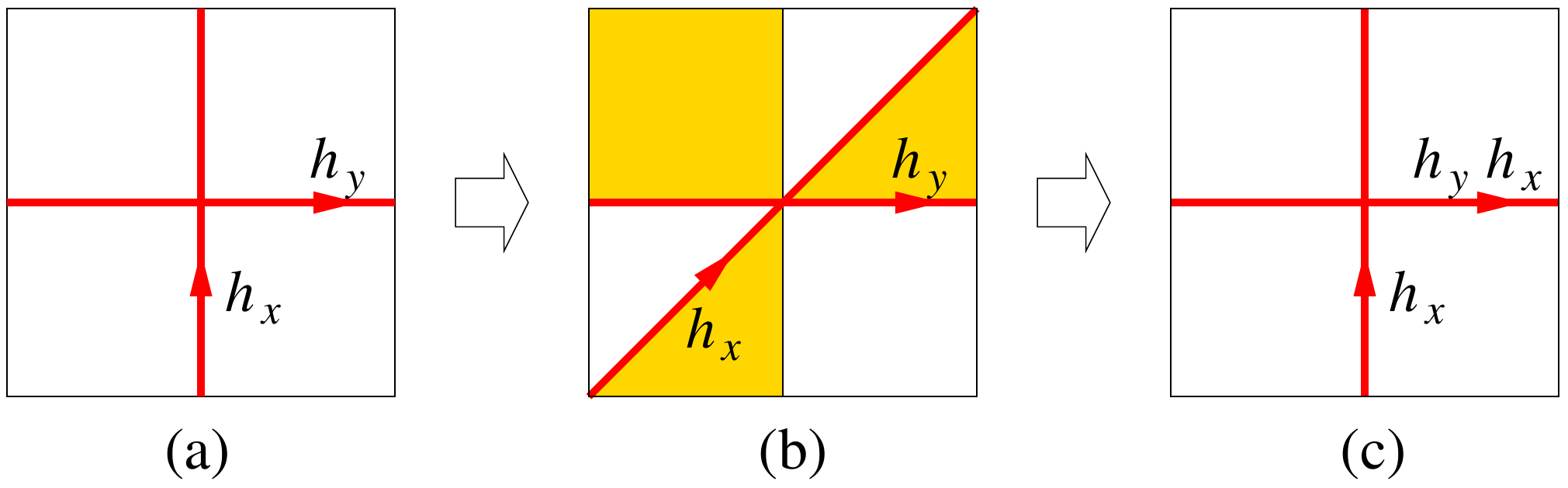}
\caption{ $\hat T$-move is the Dehn twist followed by a symmetry transformation
$h_x$ in the shaded area.  }
\label{Tmov}
\end{figure}

Note that \ml{H} and \ml{H_h} have the same energy spectrum.  So the symmetry
twist costs no energy.  Let \ml{|\Psi_{(h_x,h_y)}\>} be the ground state of
\ml{H_{h_x,h_y}} on a torus with symmetry twists \ml{h_x,h_y} in \ml{x}- and
\ml{y}-directions.  \ml{|\Psi_{(h_x,h_y)}\>} simulates the degenerate ground
states for topologically ordered phases.
We can use  \ml{|\Psi_{(h_x,h_y)}\>} to construct
$S,T$ matrices that characterize the SPT order (see Fig. \ref{Smov} and Fig. \ref{Tmov}):\\
\ml{\hat S} move:
\ml{\<\Psi_{(h_y^{-1},h_x)}|\hat{S} |\Psi_{(h_x,h_y)}\> =
 S_{h_x,h_y}\ e^{-f_S L^2+o(L^{-1})}}
\\
\ml{\hat T} move:
\ml{ \< \Psi_{(h_x,h_yh_x)}|\hat{T}|\Psi_{(h_x,h_y)}\> =
 T_{h_x,h_y}\ e^{-f_T L^2+o(L^{-1})}}
\\
\ml{\hat U} move:  \ml{\<\Psi_{(h_th_xh_t^{-1},h_th_yh_t^{-1})}|\hat{U}(h_t) |\Psi_{(h_x,h_y)}\> = U_{h_x,h_y}(h_t)}
\\
Note that in addition to the $\hat S$- and $\hat T$-moves, the SPT invariants
also contain $\hat U$-move generated by the global symmetry transformation
$h_t\in G$.

\begin{figure}[tb]
\centering
\includegraphics[height=1.3in]{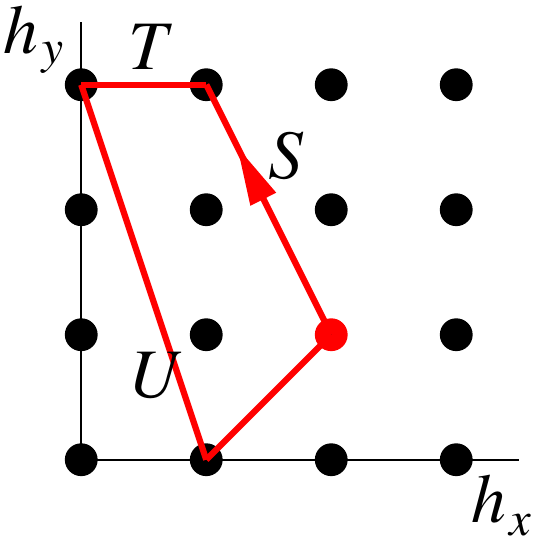}
\caption{A closed orbit in the $(h_x,h_y)$ space.}
\label{closed_orbit}
\end{figure}

The $\hat S$-, $\hat T$-, and $\hat U$-moves shift $(h_x,h_y) \to (h_x',h_y')$:
\begin{align}
 \hat S: (h_x,h_y) &\to (h_x',h_y') =(h_y^{-1},h_x);
\nonumber\\
 \hat T: (h_x,h_y) &\to (h_x',h_y') =(h_x,h_yh_x);
\nonumber\\
 \hat U(h_t): (h_x,h_y) &\to (h_x',h_y') =(h_th_xh_t^{-1},h_th_yh_t^{-1}).
\end{align}
When $(h_x',h_y')\neq (h_x,h_y)$, the complex phases
$ S_{h_x,h_y},
 T_{h_x,h_y},
 U_{h_x,h_y}(h_t)$ are not well defined, since they depend on the choices of the phases of
$|\Psi_{(h_x,h_y)}\>$ and
$|\Psi_{(h_x',h_y')}\>$.
However, the product of \ml{S_{h_x,h_y},T_{h_x,h_y},U_{h_x,h_y}(h_t)} around a
closed orbit $(h_x,h_y) \to (h_x',h_y') \to \cdots \to (h_x,h_y)$ is universal
(see Fig. \ref{closed_orbit}).  We believe that those products for various
closed orbits
%(and their generalization)
completely characterize the 2+1D SPT states.

For example, $N$ $\hat T$-moves always form a closed orbit
for Abelian $\Z_N=\{h=0,\cdots,N-1\}$ group.
For 2+1D \ml{Z_N} SPT state
labeled  by \ml{k \in H^3[Z_N,U(1)]=\Z_N}, it has one
SPT invariant:
\ma{
T_{h_xh_y^{N-1},h_y}
& \cdots
T_{h_xh_y^2,h_y}
T_{h_xh_y,h_y}
T_{h_x,h_y}
=\ee^{2\pi \ii (h_x-1)^2 k/N},
\nonumber\\
& h_x,h_y\in \Z_N.}
Such an SPT invariant completely characterizes the  2+1D \ml{Z_N} SPT state.

\subsection{Universal wavefunction overlap in 1+1D}

%\ma{O_{(h_x,h_y),(h_x',h_y')} e^{-f_O L^2+o(L^{-1})}=
%\<\Psi_{(h_x,h_y)}|\hat O |\Psi_{(h_x',h_y')}\>,\ \ \ \hat O=\hat S,\hat T,\hat U(g)}

\begin{figure}[tb]
\centering
\includegraphics[height=0.7in]{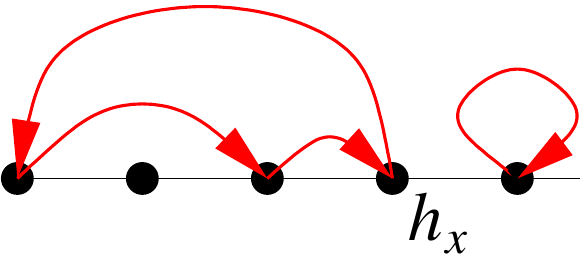}
\caption{Two closed orbits in $h_x$ space.}
\label{closed_orbit1D}
\end{figure}

In 1+1D, the SPT invariants are very simple. We only have the \ml{\hat U}-move:
\ml{\<\Psi_{(h_th_xh_t^{-1})}|\hat{U}(h_t) |\Psi_{(h_x)}\> = U_{h_x}(h_t)},
which generates the shift $h_x\to h_th_xh_t^{-1}$.  Similar to the 2+1D cases,
the product of \ml{U_{h_x}(h_t)} around a closed orbit is well defined and
universal (see Fig \ref{closed_orbit1D}).
In particular, for Abelian symmetry group,
$U_{h_x}(h_t)$ itself is universal.

\section{A 1+1D $Z_{N_1}\times Z_{N_2}$ SPT state and its Bi-kink bulk dynamical action}

\subsection{A simple example}

Now let us apply the results obtained in the last Section to a 1+1D $Z_{N_1}\times Z_{N_2}$ bosonic
SPT state, which is classified by
\ma{ H^2[Z_{N_1}\times Z_{N_2}, U(1)] =
\Z_{N_{12}}=\{0,1,\cdots, N_{12}-1\} } where
\ml{N_{12}=\text{gcd}(N_1,N_2)}.  We consider an SPT state labeled by \ml{k\in
\Z_{N_{12}}}.
% and assume \ml{N_1=N_2=N}.

%\subsection{The SPT invariant \ml{U_{h_x}(h_t)}}

\begin{figure}[tb]
\centering
\includegraphics[height=0.8in]{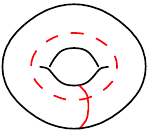}
\caption{Space-time $S^1\times S^1$ with two symmetry twists in \ml{x,t} directions.}
\label{T2symmtwst}
\end{figure}

The group elements of $Z_{N_1}\times Z_{N_2}$
are labeled by $h=(h^1,h^2),\ h^1 \in \Z_{N_1},\ h^2 \in \Z_{N_2}$.
The universal wavefunction overlap (the SPT invariant \ml{U_{h_x}(h_t)}) is
\ma{
&\ \ \ \
\<\Psi_{(h^1_x,h^2_x)}|\hat{U}(h^1_t,h^2_t) |\Psi_{(h^1_x,h^2_x)}\>
\nonumber\\
&
= U_{h^1_x,h^2_x}(h^1_t,h^2_t) =e^{ \ii k \frac{2\pi}{N_{12}}
(h^1_xh^2_t-h^2_xh^1_t)},
}
which can also be viewed as the fixed-point partition function
on space-time \ml{T^2=S^1\times S^1}
with symmetry twists in \ml{x,t} directions (see Fig \ref{T2symmtwst}):
\ma{
\label{SPTinv1D}
Z_\text{fixed-point} = U_{h^1_x,h^2_x}(h^1_t,h^2_t) =
e^{ \ii k \frac{2\pi}{N_{12}}
(h^1_xh^2_t-h^2_xh^1_t)}}

Both wavefunction overlap and partition function pictures imply
the following physical meaning for the above SPT invariant:
a symmetry twist of \ml{Z_{N_1}} carries \ml{Z_{N_2}}-charge \ml{k}.
\ma{
\<\Psi_{(h^1_x,h^2_x)=(1,0)}|\hat{U}(h^1_t=0,h^2_t=1) |\Psi_{(h^1_x,h^2_x)=(1,0)}\>
= e^{ \ii k \frac{2\pi}{N_{12}} },
}

%\subsection{An example}

Let us discuss a concrete example for the above 1+1D SPT invariant.  We
consider a spin-1 chain with the spin-rotation symmetry \ml{Z_2\times Z_2=D_2} = \ml{180^\circ}
 in \ml{S^x,S^z}.
The Hamiltonian on a ring is given by (untwisted case)
\ma{
\label{Jxyz}
H_{D_2}  &= \sum_{i=1}^{L-1} (J_x S^x_iS^x_{i+1} +J_y S^y_iS^y_{i+1} +J_z S^z_iS^z_{i+1})
\nonumber\\
&\ \ \ \
+J_x S^x_LS^x_{1} +J_y S^y_LS^y_{1} +J_z S^z_LS^z_{1}
}
where $J_x=J_y=J_z>0$.  The ground state carries a trivial quantum number
\ml{e^{i \pi \sum S^z_i} } with \ml{e^{i \pi \sum S^z_i} =1}.

If we add a
twist by \ml{e^{i\pi \sum S^x_i}}, the Hamiltonian becomes
\ma{
H^\text{twist}_{D_2}  &= \sum_{i=1}^{L-1}
(J_x S^x_iS^x_{i+1} +J_y S^y_iS^y_{i+1} +J_z S^z_iS^z_{i+1})
\nonumber\\
&\ \ \ \
+J_x S^x_LS^x_{1} -J_y S^y_LS^y_{1} -J_z S^z_LS^z_{1}
}
The twisted ground state carries a non-trivial quantum number
\ml{e^{i \pi \sum S^z_i} } with \ml{e^{i \pi \sum S^z_i} =-1}.
Such a dependence of the ground state  quantum number
\ml{e^{i \pi \sum S^z_i} } on the  \ml{e^{i\pi \sum S^x_i}} twist is the
1+1D SPT invariant discussed above.

%\subsection{From a SPT invariant to a SPT mechanism}

The above SPT invariant also suggests a  mechanism for the 1+1D
\ml{Z_{N_1}\times Z_{N_2}} SPT state.  We notice that the SPT invariant implies
a symmetry twist of \ml{Z_{N_1}} that carries a ``charge'' of  \ml{Z_{N_2}}.  Since
the symmetry twist of \ml{Z_{N_1}} is the domain wall of \ml{Z_{N_1}} in a
\ml{Z_{N_1}} symmetry breaking state, we may (1) start with a \ml{Z_{N_1}}
symmetry breaking state, (2) bind \ml{k} \ml{Z_{N_2}}-charge to the domain wall
of \ml{Z_{N_1}}, and (3) restore the  \ml{Z_{N_1}} symmetry by proliferating the
domain walls. In this way, we obtain a 1+1D \ml{Z_{N_1}\times Z_{N_2}} SPT state
labeled by \ml{k\in \cH^2[Z_{N_1}\times Z_{N_2}, U(1)]}.

\begin{figure}[tb]
\centering
\includegraphics[height=0.4in]{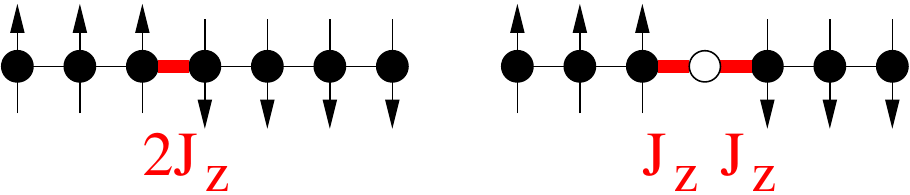}
\caption{Two kinds of domain walls with the same energy, but different
\ml{Z_2^z}-charges, $0$ (mod 2) and $1$ (mod 2) respectively on a lattice.
Eq.(\ref{eq:hopping1})'s $H^\text{hop}_{1}$ is a hopping operator for the first kind of domain wall.
Eq.(\ref{eq:hopping2})'s $H^\text{hop}_{2}$ is a hopping operator for the second kind of domain wall.
}
\label{dwall}
\end{figure}

For example, let us consider a 1D \ml{Z_2^x\times Z_2^z} spin-1 chain
with symmetry
\begin{align}
Z_2^x:\ & (|\up_x\>,|0_x\>,|\down_x\>)\to (|\up_x\>,-|0_x\>,|\down_x\>)
\nonumber\\
Z_2^z:\ & (|\up_z\>,|0_z\>,|\down_z\>)\to (|\up_z\>,-|0_z\>,|\down_z\>)
.
\end{align}
The following Hamiltonian has the \ml{Z_2^x\times Z_2^z} symmetry
\ma{
H^0_{Z_2\times Z_2}  = \sum_i -J_z S^z_iS^z_{i+1}
}
but its ground state breaks the \ml{Z_2^x} symmetry.  Such a symmetry breaking
state has two kinds of domain walls which happen to have the same energy,  but
different \ml{Z_2^z}-charges.  The two kinds of domain walls, shown in Fig.\ref{dwall}, have different
hopping operators:
\begin{eqnarray}
\label{eq:hopping1}
H^\text{hop}_{1}  &= \sum_i -\frac{K}{2} \big( (S^+_i)^2 +(S^-_i)^2 \big) \nonumber\\
&= \sum_i -K \big( (S^x_i)^2-(S^y_i)^2 \big),
\end{eqnarray}
\begin{eqnarray}
\label{eq:hopping2}
H^\text{hop}_{2}  &= -\sum_i \frac{J_{xy}}{2}(S^+_i S^+_{i+1} + S^-_iS^-_{i+1}) \nonumber\\
&= \sum_i J_{xy}(-S^x_iS^x_{i+1} + S^y_iS^y_{i+1}) .
\end{eqnarray}
%\textbf{the first term $-S^x_iS^x_{i+1}$ has a minus sign.)}
Here we used the fact that $S^+_i \equiv  S^x_i+ \ii S^y_i$ and $S^-_i \equiv  S^x_i- \ii S^y_i$.
It is straightforward to see the $(S^+_i)^2$ operator hops the first kind of domain wall of Fig.\ref{dwall} in one direction, while
the $(S^-_i)^2$ operator hops the first kind of domain wall of Fig.\ref{dwall} in the opposite direction.
On the other hand, the $S^+_i S^+_{i+1}$ operator hops the second kind of domain wall of Fig.\ref{dwall} in one direction, while
the $S^-_iS^-_{i+1}$ operator hops the second kind of domain wall of Fig.\ref{dwall} in the opposite direction.

Adding a strong enough hopping operator can make a domain wall subject to a
negative energy cost, which restores the  \ml{Z_2^x} symmetry by proliferating
the domain walls.  We find that \ml{H^0_{Z_2\times Z_2}+H^\text{hop}_{1}} leads
to a trivial SPT state, while \ml{H^0_{Z_2\times Z_2}+H^\text{hop}_{2}} leads
to a non-trivial \ml{Z_2\times Z_2} SPT state. Via a unitary transformation, the Hamiltonian \ml{H^0_{Z_2\times Z_2}+H^\text{hop}_{2}} is equivalent to
the Hamiltonian of Eq.\eq{Jxyz} discussed above, as the Haldane phase of a spin-1 anti-ferromagnetic Heisenberg chain.

%topological  %Move bulk dynamical effective action to Appendix
\subsection{Bi-kink topological term NL$\sigma$M and dynamic gauge theory}
The underlying physics of the above 1+1D $Z_{N_1}\times Z_{N_2}$ SPT state can also be captured by the following Higgs action with a \emph{bi-kink topological term}:
\begin{eqnarray}\label{bi-kink}
\mathcal{L}_\text{bi-kink}&=&{\chi\over2} (\partial_\mu \theta^I)^2+\frac{\ii}{2}C^{IJ}\varepsilon^{\mu\nu}
\partial_\mu \theta^I\partial_\nu \theta^J\nonumber\\
&\simeq&{\chi\over2} (\partial_\mu \theta^I_{\s}+b_\mu^I)^2\\&+&\frac{\ii}{2}C^{IJ}\varepsilon^{\mu\nu}
(\partial_\mu \theta^I_{\s} + b_\mu^I)(\partial_\nu \theta^J_{\s} + b_\nu^J)+\mathcal{L}_\text{Maxwell}^b,\nonumber
\end{eqnarray}
where $I=1,2$ and the structure constant $C_{IJ}$ is totally antisymmetric
with $C_{IJ}=-C_{JI}$. We assume Einstein summations for repeated indices throughout the whole paper.
\cblue{The quantum phase fluctuation can be captured by a real scalar compact field $\theta^I\equiv\theta^I_{\text{s}}+\theta^I_{\text{v}}$ with a smooth piece and a singular piece $\theta^I_{\text{s}}$ and $\theta^I_{\text{v}}$.
%To achieve the superfluid state, we condense the boson (whose boson field operator has a $U(1)$ phase $e^{\ii \theta}$),
%such that the quantum phase is coherent to an almost fixed value, see Appendix \ref{sec:disordersf}.
%On the other hand,
To achieve the disordered insulator state, we can condense the vortex, namely strongly disorder the superfluid coherent phase. We will write
$\partial_\mu \theta^I_{\text{s}}+\partial_\mu \theta^I_{\text{v}} \equiv \partial_\mu \theta^I_{\s} +b_\mu^I$.
%as we derived in Eq.(\ref{eq:singular_b}).
The $\partial_\mu \theta^I_{\s}$ captures
the smooth piece $\partial_\mu \theta^I_{\text{s}}$, and the additional $b_\mu^I$ captures the singular piece $\partial_\mu \theta^I_{\text{v}}$.}
We note that the real scalar fields $\theta^I_\s$ can be viewed as the phase fluctuations of $Z_{N_I}$ symmetry in a $Z_{N_1}\times Z_{N_2}$ symmetry breaking phase while vector fields $b_\mu^I$ (with $\mathcal{L}_\text{Maxwell}^b$ the corresponding Maxwell term) describe the proliferations of domain walls, which restore the $Z_{N_1}\times Z_{N_2}$ symmetry.
%Actually the bi-kink topological term origins from a current-current topological term:
%\begin{eqnarray}
%\frac{\ii}{2}C^{IJ}\varepsilon^{\mu\nu}j_\mu^I j_\nu ^J,
%\end{eqnarray}
%where $j^I_\mu=i({\phi^I}^*\partial_\mu \phi^I-\phi^I\partial_\mu {\phi^I}^*)$ and $\phi^I$ is corresponding complex fields for $Z_N^I$ charge bosons.) In the Higgs phase, $\phi^I=\sqrt{\rho^I}e^{i\theta^I}$
Such a Higgs action with a bi-kink topological term will enforce a \ml{Z_{N_1}} domain wall that carries a ``charge'' of  \ml{Z_{N_2}}, and vice versa.
It is clear that the bi-kink topological
term is just a boundary term in the absence of gauge fields $b_\mu^I$. In the following we will show that such a bulk action \cblue{with the bi-kink topological term} indeed describes
the $Z_{N_1}\times Z_{N_2}$ SPT physics in 1+1D.

After dropping the total derivative term, we
can rewrite the above action as:
\begin{eqnarray}
\mathcal{L}_\text{bi-kink}&=&{\chi\over2} (\partial_\mu \theta^I_\s+b_\mu^I)^2\\&+&\frac{\ii}{2}C^{IJ}\varepsilon^{\mu\nu}
(-2\theta^I_\s\partial_\mu  b_\nu^J + b_\mu^I b_\nu^J)+\mathcal{L}_\text{Maxwell}^b,\nonumber
\end{eqnarray}
Next, we introduce the Hubbard-Stratonovich fields $j^\mu_I$ to decouple the quadratic term as
\begin{eqnarray}
\mathcal{L}_\text{bi-kink}&=&\frac{1}{2\chi}(j^\mu_I)^2 \cblue{-\ii} \theta^I_\s \partial_\mu j^\mu_I + \ii  b_\mu^I j^\mu_I \nonumber\\&+&
\frac{\ii}{2}C^{IJ}\varepsilon^{\mu\nu} (-2\theta^I_\s \partial_\mu  b_\nu^J + b_\mu^I b_\nu^J) + \mathcal{L}_\text{Maxwell}^b ,\nonumber
\end{eqnarray}
Integrating out the smooth fluctuations $\theta_\s^I$ leads to the following constraint:
\cblue{
\begin{eqnarray}
\partial_\mu(j^\mu_I+C^{IJ}\varepsilon^{\mu\nu}  b_\nu^J )=0.
\end{eqnarray}
}
The above constraints can be solved by:
\cblue{
\begin{eqnarray}
j^\mu_I=\frac{1}{2\pi}\varepsilon^{\mu\nu}\partial_\nu a^I- C^{IJ}\varepsilon^{\mu\nu}  b_\nu^J.
\end{eqnarray}
}
where $a^I$ do not need to be globally defined.
%NOT In the London limit
To disorder the $U(1)$ phase, we take $\chi \ll \chi_c$, we can drop out the $\frac{1}{2\chi}(j^\mu_I)^2$ term as well as the Maxwell term of
gauge fields $b_\mu^I$ thanks to their RG irrelevancy \cite{Wenbook}. We end up with an effective topological action:
\begin{eqnarray} \label{eq:1+1DSPT}
\mathcal{L}_\text{top}=\frac{\ii}{2\pi}\varepsilon^{\mu\nu}  \cblue{b_\mu^I  \partial_\nu a^I} +\cblue{\frac{-\ii}{2}}C^{IJ}\varepsilon^{\mu\nu}b_\mu^I b_\nu^J,
\end{eqnarray}
%
%\cblue{{\bf (Juven: I modify some factor $\pm $ for $\ii$. We are doing Euclidean.
%There may be also a singular term $\theta \partial_\mu \partial_\nu \theta =bb$.)}}
%
%
The gauge transformation of $b_\mu^I$ in the above action will induce a shift on the \cblue{scalar} fields $a^I$:
\begin{eqnarray}
a^I \rightarrow a^I + 2\pi C^{IJ}g^J;\quad b_\mu^I\rightarrow b_\mu^I+\partial_\mu g^I.\label{gauge}
\end{eqnarray}
The above functions do not necessarily need to be \emph{globally defined}. In fact, the compactness condition of $a^I$ and $b_\mu^I$
%should be compact and they are defined subject to the following constraints:
%\emph{Globally defined}
%means that the function is continuous and differentiable on the whole manifold.
%We can assign the gauge field fiber everywhere globally on a based manifold
%and we do not need \emph{local charts} to pave the base-manifold.
%Here $f^I$ needs to be $2 \pi \times\,$integer in order to have $\ee^{\ii a^I}$ globally defined,
%though $a^I$ may not be a globally defined real function.
%But $\ee^{\ii a^I}$ must be globally defined.
%Because of the $\dd (g^I \dd g^J)$ boundary term needs to be interated as zero on a closed manifold,
%we also require $g^I$ to be globally defined. Thus $\ee^{\ii g^I}$ must be
%globally defined.  Such gauge transformation suggests that $b^I$ is globally defined.
%In short, $\ee^{\ii a^I}$, $g^I$, and $b^I$ are globally defined. However,
%$a^I$ need not to be globally defined.
%The globally defined or not of the gauge field
implies the closed loop or the closed surface integral has the constraints:
\begin{align}
\oint \dd a/(2\pi)  \in \mathbb{Z},\;\;\;
%\ \ \ \ \
\Ointint \dd b_I/(2\pi)  \in \mathbb{Z}.
\end{align}
In Sec.\ref{sec:partition-GSD}, we will derive the same constraints in the path integral level,
from the constraints of $U(1)$ charge and the vortex number on a closed-surface.

% wrong version:
%Interestingly, the gauge transformation of $b_\mu^I$ in the above action will induce a shift on the \cblue{scalar} fields $a^I$:
%\begin{eqnarray}
%a^I \rightarrow a^I \cblue{+} 2\pi C^{IJ}g^J;\quad b_\mu^I\rightarrow b_\mu^I+\partial_\mu g^I
%\end{eqnarray}
%where $g^I$ do not need to be globally defined by $\ee^{\ii g^I}$ must be
%globally defined.  Such gauge transformation suggest that $b^I$ is a compact
%$U(1)$ gauge field (one form).  Also $a^I$ is not a globally defined real
%function. But $\ee^{\ii a^I}$ must be globally defined.

Now, let us compute the quantization condition for the coefficients $C_{IJ}$.
We note that the average of
$\th^I=\th^I_\s+\th_\text{v}^I$ is quantized as $2\pi/N_I \times$ integer. In the disordered phase
which restores the $Z_{N_1}\times Z_{N_2}$ symmetry, $\th^I$'s
have many fluctuating kinks.  Let us consider a
configuration where $\th^1$ has a kink $\Del \th^1=2\pi k_1/N_1$ on the $t$
axis and $\th^2$ has a kink $\Del \th^2=2\pi k_2/N_2$ on the
$x$ axis.  For such a configuration, the action from the bi-kink
topological term is given by
\begin{align}
 S&=\int \dd x\dd t\; \frac{\ii}{2} C_{IJ}\varepsilon^{\mu\nu}
\partial_\mu \theta^I\partial_\nu \theta^J
\nonumber\\
& =8\pi^2 \ii C_{12} \frac{k_1k_2}{N_1N_2}.
\end{align}
%\cblue{{\bf (what is the relation between $\chi$ and $C_{12}$ then?)}}
This means that the $\th^1$ kink carries a
$Z_{N_2}$-charge $2\pi  C_{12} \frac{k_1}{N_1}$ mod $N_2$.
Since $k_1=0 \sim k_1=N_1$, $C_{12}$ must be quantized:
\begin{align}
 2\pi  C_{12}=0 \text{ mod } N_2,
\ \
 2\pi  C_{12}=0 \text{ mod } N_1. \label{quantization}
\end{align}
Thus
\begin{align}
  C_{12}=\frac{p_\text{II}}{2\pi} \frac{N_1N_2}{N_{12}}, \ \ \ \ p_\text{II}=0,\cdots,N_{12}-1 \label{value}
\end{align}
where $N_{12}=\text{gcd}(N_1,N_2)$.  Also we note that $C_{12}$ has only
$N_{12}$ distinct quantized values, corresponding to $N_{12}$ distinct charge
assignments.

The above argument for the quantization condition of $C_{12}$ due to global $Z_{N_1}\times Z_{N_2}$ symmetry can also be derived in a rigorous way
by adding a coupling term to external background gauge field $A^I$:
%To probe the $Z_{N_1}\times Z_{N_2}$ SPT states of the above bulk dynamical action, let us introduce external gauge fields $A^1$ and $A^2$ to couple to the above bulk action with:
\begin{eqnarray}
\mathcal{L}_\text{coupling}=\frac{\ii}{2\pi}\varepsilon^{\mu\nu}A^I_\mu \partial_\nu a^I,
\end{eqnarray}
As the physical meanings of
$A^1$ and $A^2$ are $Z_{N_1}$ and $Z_{N_2}$ symmetry twists, $A^I$ must be a flat connection with $\dd A^I =0$ and $\oint A^I=2\pi n_I/N_I$. On the other hand, since $\int \dd x \dd t \mathcal{L}_\text{coupling}$ must be invariant under gauge transformation Eq.(\ref{gauge}), $C_{12}$ can not take arbitrary value.
A shot calculation suggests the same quantization condition Eq.(\ref{quantization}).
%we can introduce a Lagrangian multiplier term $\frac{\ii \cblue{N_I}}{2\pi}\varepsilon^{\mu\nu} B^I \partial_\nu A^I_\mu$ to implement such a condition on $\varepsilon^{\mu\nu}  \partial_\nu A^I_\mu=0$.

%To well-define the field theory, we not only need to know the \emph{Lagrangian}, but also need to define the \emph{path integral partition function}
%and the \emph{field constraints}.

In Sec.\ref{sec:partition-GSD}, we will define a rigorous SPT internal gauge theory path integral, and
we confirm that the GSD of our theory is unique on a closed manifold, GSD=1,in agreement with SPT state.
We will also derive the \emph{SPT invariant} in Ref.\cite{JuvenSPT1} by coupling the internal gauge theory to semi-classical probed field $A$.
In this way, it becomes manifested that $C_{12}$ can only take $N_{12}$ distinguishable value derived in Eq.(\ref{quantization}).
In the following, we %will try to
generalize the above results to higher dimensions.

\section{A review of Chern-Simons action approach to $2+1$D Abelian SPT states}
In this section, we will start with a brief review on the Chern-Simons action
approach for 2+1D Abelian SPT states. Then we explain the physical meaning of the Chern-Simons action approach and discuss its limitations.

It is well known that a vortex condensation can turn a boson
superfluid into a trivial bosonic insulator.  A bosonic $U(1)$ SPT state is
also a  bosonic insulator, but a non-trivial one.  It turns out that a
condensation of vortex-charge bound state can turn  a boson superfluid into a
non-trivial $U(1)$ SPT state.

To show this,
let us consider a boson superfluid for one specie of bosons, which can be
described by an XY model:
\begin{eqnarray}\label{XY1}
\mathcal{L}_\text{XY} = {1\over2} (\partial_\mu \theta)^2,
\end{eqnarray}
If the vortex of the boson condenses, $\theta$ in the XY
model is no longer a smooth function of space-time. We can introduce the
singular part by replacing $\partial_\mu\theta$ by $\partial_\mu\theta_\s+b_\mu$,
where the field strength of gauge field $b_\mu$ corresponds to the vortex
current density $\tilde J^\mu={1\over
2\pi}\varepsilon^{\mu\nu\lambda}\partial_\nu b_\lambda$.

The charge of gauge field $b_\mu$ is the number of vortices minus the number of
anti-vortices and is quantized.  In the vortex condensed phase, the phase
fluctuation of the vortex condensate can be described by another XY model,
which is dual to the Maxwell term of the gauge field $b_\mu$. Now the boson
superfluid is described by the following  Lagrangian
\begin{eqnarray}\label{XY2}
\mathcal{L}_\text{Higgs}={1\over2}[(\partial_\mu \theta_\s + b_\mu)^2 + {1\over4\pi^2}\tilde F_{\mu\nu}\tilde F^{\mu\nu}],
\end{eqnarray}
where $\tilde F_{\mu\nu}=\partial_\mu b_\nu-\partial_\nu b_\mu$ and we have normalized with $v=1, \chi=1$.

We can introduce a Hubbard-Stratonovich field $j_\mu$ to decouple the quadratic term as
\begin{eqnarray}
\mathcal{L}_\text{Higgs}=\frac{1}{2}(j^\mu)^2-\ii \theta_\s \partial_\mu j^\mu+ \ii b_\mu j^\mu + {1\over8\pi^2}\tilde F_{\mu\nu}\tilde F^{\mu\nu}.\nonumber
\end{eqnarray}
Integrating out the $\theta_\s$ field results in a constraint $\partial_\mu j^\mu=0$. From this constraint, we can write $j^\mu={1\over2\pi}\varepsilon^{\mu\nu\lambda}\partial_\nu a_\lambda$. %Similar to $\tilde a_\mu$,
The charge of $a_\mu$ is equal to the boson number and is quantized. %non-compact since instantons are not allowed.
With these results, the path integral becomes
\begin{eqnarray}
\mathcal{L}_\text{BF}&=&{\ii \over2\pi}\varepsilon^{\mu\nu\lambda}b_\mu\partial_\nu a_\lambda+{1\over8\pi^2}[\tilde F_{\mu\nu}\tilde F^{\mu\nu}+F_{\mu\nu}F^{\mu\nu}],\;\;\;\;\label{CS}
\end{eqnarray}
where $F_{\mu\nu}=\partial_\mu a_\nu-\partial_\nu a_\mu$.  Note that the boson
current $j^b_\mu = {1\over2\pi}\varepsilon^{\mu\nu\lambda}\partial_\nu
b_\lambda$, while the vortex current $j^v_\mu =
{1\over2\pi}\varepsilon^{\mu\nu\lambda}\partial_\nu a_\lambda$.

The above can be generalized to the case with $k$-species of bosons with
$U^k(1)$ symmetry.  The bosonic insulator induced by the vortex condensation is
described by the following Chern-Simons action:
\begin{equation}
\mathcal{L}_\text{CS}=\frac{\ii }{4\pi}\varepsilon^{\mu\nu\lambda}K_{IJ}^0a_\mu^{I} \partial_\nu a_\lambda^{J};\quad I=1,2,\cdots,2k \label{CS}
\end{equation}
with
\begin{equation}
\label{canonical}{K^0}_{IJ}=\left(
                \begin{array}{cc}
                  0 & 1 \\
                  1 & 0 \\
                \end{array}
              \right)\otimes \textbf{I}_{k\times k}
\end{equation}
where $a^{2k}_\mu \sim a_\mu$ and $a^{2k-1}_\mu \sim b_\mu$.  Since
$|\det[K]|=1$, the above Chern-Simons action has a unique ground state %degenerate
$1$ on any closed manifold. The chiral central charge for the edge states is given by
the signature of $K$ which is zero. So the bosonic insulator has a trivial
topological order.

%One way to understand the classification of Abelian SPT states based on above
%effective field theory is to couple the system to several $U(1)$ external gauge
%field.  The minimal coupling between the matter fields in the bulk and the
%external gauge fields $A^\alpha$ reads:

However, the bosonic insulator may have a non-trivial $U^k(1)$ SPT order.
To see this, we turn on the external $U^k(1)$ gauge field
$A^\alpha_\mu$ to reveal the  $U^k(1)$ symmetry of the theory:
\begin{equation}
  \mathcal{L}_\text{coupling}= \frac{\ii }{2\pi}\varepsilon^{\mu\nu\lambda}
q_{\alpha}^I {A}^\alpha_{\mu} \partial_\nu a^I_\lambda;\quad \alpha=1,2,\cdots,k \label{coupling}
\end{equation}
Here $\v q_{\alpha}$ are integer-value charge vectors.  $q^{2l-1}_\alpha$
is the $A^\alpha$-charge carried by the $l^\text{th}$-species of bosons,
and $q^{2\beta}_\alpha$ is the $A^\alpha$-charge carried by the vortex of the
$l^\text{th}$-species of bosons.  We see that charge vectors $\v
q_{\alpha}$ \cblue{reveal} the information on what kinds of vortex-charge bound states are
condensing to produce the bosonic insulator.  Different vortex-charge bound
states (\ie different  charge vectors) will lead to different $U^k(1)$ SPT
orders.

The full theory is given by
$\mathcal{L}=\mathcal{L}_\text{CS}+\mathcal{L}_\text{coupling}
%+\mathcal{L}_\text{Higgs}[\varphi_\alpha, A_\alpha]
$,
%where the Higgs part break the $U(1)$'s to $Z_{N_i}$'s.  We then
%substitute $j^{\mu}_I=\frac{1}{2\pi}\varepsilon^{\mu\nu\lambda}\partial_\nu
%a_\lambda^I$ and
After integrating out internal gauge fields $a_\mu^I$ (the matter fields), we obtain an
effective theory for the external fields $A^\alpha$:
\begin{equation}
  \mathcal{L}_\text{eff}=\cblue{-} \frac{\ii }{4\pi}\varepsilon^{\mu\nu\lambda}{A}^\alpha_\mu {q}_\alpha^I{K^0}_{IJ}{q}_\beta^J\partial_\nu {A}_\lambda^\beta.
%+\mathcal{L}_\text{Higgs}[\varphi_\alpha, A_\alpha].
  \label{gaugedL}
\end{equation}
By considering equivalent class of response $K$ matrix
$\widetilde{K}_{\alpha\beta}\equiv {q}_\alpha^I{K^0}_{IJ}{q}_\beta^J$, we can
``classify'' 2+1D $U^k(1)$ SPT states described by the Chern-Simons theory
Eq.(\ref{CS}).
%, or its equivalent $BF$ theory Eq.(\ref{BF}).
We can also break the $U^k(1)$ symmetry down to $Z_{N_1}\times \cdots\times
Z_{N_k}$ symmetry and obtain a ``classification'' of  $Z_{N_1}\times
\cdots\times Z_{N_k}$ SPT states in 2+1D.
Since $Z_{N_\al}$ group can always be embedded into $U^k(1)$ group, it is
not a surprise that the $Z_{N_1}\times \cdots\times Z_{N_k}$  SPT state can
be described by the same Chern-Simons action.

However, since $H^3[Z_{N_1}\times \cdots \times Z_{N_k} , U(1)]=\oplus_{i}\Z_{N_{i}}\oplus_{i<j}\Z_{N_{ij}}\oplus_{i<j<k}\Z_{N_{ijk}}$( $N_{ijk}=\text{gcd}(N_i,N_j,N_k)$),
the above classification turns out to be incomplete and it can only
describe a subclass of Abelian SPT states labeled by $\oplus_{i}\Z_{N_{i}}\oplus_{i<j}\Z_{N_{ij}}$, namely, the type I and type II SPT
phases. In the following, we will develop an effective field theory description
for type-III SPT order in 2+1D, which is labeled by $\oplus_{i<j<k}\Z_{N_{ijk}}$.

\section{A 2+1D $Z_{N_1}\times Z_{N_2}\times Z_{N_3}$ SPT state and its Tri-kink bulk dynamical action}
\subsection{A 2+1D $Z_{N_1}\times Z_{N_2}\times Z_{N_3}$ SPT state }
Without loss of generality, it is sufficient to discuss a 2+1D $Z_{N_1}\times Z_{N_2}\times Z_{N_3}$
bosonic SPT state, which is classified by
\ma{
&\ \ \ \
H^3[Z_{N_1}\times Z_{N_2}\times Z_{N_3} , U(1)]
\\
&
=
\Z_{N_1}\times
\Z_{N_2}\oplus
\Z_{N_3}\oplus
\Z_{N_{12}}\oplus
\Z_{N_{23}}\oplus
\Z_{N_{13}}\oplus
\Z_{N_{123}}
\nonumber
}
We consider a type III SPT state labeled by \ml{k\in  \Z_{N_{123}}}
%and assume \ml{N_1=N_2=N_3=N}.

The group elemenets of $Z_{N_1}\times Z_{N_2}\times Z_{N_3}$
are labeled by $h=(h^1,h^2,h^3),\ h^1 \in \Z_{N_1},\ h^2 \in \Z_{N_2},\ h^3 \in \Z_{N_3}$.
The SPT invariant
\ml{U_{h_x,h_y}(h_t)} for the above SPT state
is the fixed-point partition function
on space-time \ml{T^3=(S^1)^3}
with symmetry twists in \ml{x,y,t} directions:
\ma{
\label{SPTinv2D}
Z_\text{fixed-point} = U_{h_x,h_y}(h_t) = e^{ \ii k \frac{2\pi}{N_{123}}
\eps_{abc} h^a_x h^b_y h^c_t }
}
The physical meaning of the SPT invariant is the following: Consider the ground
state of the Hamiltonian with symmetry twists in \ml{Z_{N_1}} and \ml{Z_{N_2}},
the intersection of the symmetry twist in \ml{Z_{N_1}} and the symmetry twist
in \ml{Z_{N_2}} carries \ml{Z_{N_3}}-charge \ml{k}.

%The above  SPT invariant leads to a mechanism for the SPT state: (1) start with
%a \ml{Z_{N_1}\times Z_{N_2}} symmetry breaking state, (2) bind \ml{k}
%\ml{Z_{N_3}}-charge to the  intersection of the domain wall of \ml{Z_{N_1}} and
%the domain wall of \ml{Z_{N_2}}, and (3) restore the  \ml{Z_{N_1}\times
%Z_{N_2}} symmetry by proliferating the domain walls. This way we obtain a 2+1D
%\ml{Z_{N_1}\times Z_{N_2}\times Z_{N_3}} SPT state labeled by \ml{k\in
%\Z_{N_{1213}}}.

The above SPT invariant also allows us to calculate the
dimension reduction of the
2+1D SPT state to a 1+1D SPT state: We view the space-time as
\ml{T^3 = T^2_{x,t}\times S^1_y}, and put $Z_{N_3}$ symmetry twist
\ml{(h^1_y,h^2_y,h^3_y)=(0,0,1)} in the small circle $S^1_y$.
The 2+1D partition function reduces to a 1+1D partition function
\ma{
Z_\text{fixed-point} = e^{ \ii k 2\pi (h^1_xh^2_t-h^2_xh^1_t)}
}
which is the SPT invariant of a 1+1D SPT state.  We find that the resulting
1+1D SPT state is the one labeled by \ml{k\in H^2[Z_{N_1}\times Z_{N_2},
U(1)]=\Z_{N_{12}}}.  The boundary of such a 1+1D SPT state carries degenerated
states that form a projective representation of \ml{Z_{N_1}\times Z_{N_2}}.
This leads to an experimental probe of the $Z_{N_1}\times Z_{N_2}\times
Z_{N_3}$ SPT state: a \ml{Z_{N_3}} ``vortex'' (end of \ml{Z_{N_3}} symmetry
twist) carries degenerated states that form a projective representation of
\ml{Z_{N_1}\times Z_{N_2}}.

The result of the above dimension reduction can also be viewed as each
$Z_{N_3}$ twist (which is a 1D curve in 2D space) carries a 1+1D
\ml{Z_{N_1}\times Z_{N_2}}  SPT state labeled by \cblue{\ml{k\in H^2[Z_{N_1}\times
Z_{N_2}},U(1)]}. This picture leads to another mechanism for the 2+1D \ml{Z_{N_1}\times
Z_{N_2}\times Z_{N_3}} SPT state: (1) start with a \ml{Z_{N_3}} symmetry
breaking state, (2) bind a 1+1D \ml{Z_{N_1}\times Z_{N_2}} SPT state to the
domain wall of \ml{Z_{N_3}}, and (3) restore the  \ml{Z_{N_3}} symmetry by
proliferating the domain walls. In this way, we obtain a 2+1D \ml{Z_{N_1}\times
Z_{N_2}\times Z_{N_3}} SPT state labeled by \cblue{\ml{k\in \Z_{N_{123}}}}.

The 2+1D SPT invariant \Eq{SPTinv2D} on space-time \ml{T^3=(S^1)^3}
can also be expressed as a topological term of probe fields $A_I$:
\ma{
\label{SPTinv2DF}
Z^\text{twist}_\text{fixed-point}(T^3) = e^{ \ii p_\text{III} \frac{N_1N_2N_3}{(2\pi)^2 N_{123}} \int A_1\wedge A_2\wedge A_3 }, \ \ \dd A_I=0,
}
with an integer $p_\text{III}$.
Again, since $A_I$ describes symmetry twists on the boundary, it must be flat connection with $\dd A_I=0$. \ml{\int A_1\wedge A_2\wedge A_3} is also gauge invariant if \ml{\dd A_I=0}.  The field
theory representation of the SPT invariants \Eq{SPTinv2DF}, should be valid for any space-time topologies. In the following we will show how to derive such a topological response from a bulk dynamical effective action.

\subsection{Tri-kink topological term NL$\sigma$M} \label{sec:Tri-kink topological}
To describe the so-called type-III $Z_{N_1}\times Z_{N_2}\times Z_{N_3}$ SPT
orders in $2+1$D, we consider the following effective action for three species
of bosons with vortex condensation.  The action contains a new \emph{tri-kink
topological term} -- the $C_{IJK}$-term (the following is a generalization of
\Eq{XY2}):
\begin{eqnarray}\label{tri-kink}
&&\mathcal{L}_\text{tri-kink}={1\over2}(\partial_\mu \theta^I)^2+\frac{\ii }{3} C_{IJK}\varepsilon^{\mu\nu\lambda}\partial_\mu \theta^I\partial_\nu \theta^J
\partial_\lambda \theta^K\nonumber\\
&\simeq&{1\over2}(\partial_\mu \theta^I_\s + b_\mu^I)^2 + \mathcal{L}_\text{\cblue{Maxwell}}^b
\\
&
+&\frac{\ii }{3} C_{IJK}\varepsilon^{\mu\nu\lambda}
(\partial_\mu \theta^I_\s+ b_\mu^I)(\partial_\nu \theta^J_\s + b_\nu^J)
(\partial_\lambda \theta^K_\s + b_\lambda^K)
\nonumber
\end{eqnarray}
where $I=1,2,3$ and the structure constant $C_{IJK}$ is totally antisymmetric
with $C_{IJK}=-C_{JIK}=-C_{IKJ}$.  It is clear that the tri-kink topological
term is just a boundary term in the absence of gauge fields $b_\mu^I$.

To understand the physical meaning of the tri-kink topological term, we first
note that the type-III SPT orders in $2+1$D only exist for \cblue{a} finite group
$Z_{N_1}\times Z_{N_2}\times Z_{N_3}$.  So we need to break the $U(1)^3$
symmetry down to $Z_{N_1}\times Z_{N_2}\times Z_{N_3}$ symmetry.  The average of
$\th^I$ is quantized as $2\pi/N_I \times$ integer.  In the disordered phase
which restores the  $Z_{N_1}\times Z_{N_2}\times Z_{N_3}$ symmetry,  $\th^I$'s
have many fluctuating kinks along space-time surfaces.  Let us consider a
configuration in the space-time where $\th^1$ has a kink $\Del \th^1=2\pi k_1/N_1$ on the $y$-$t$
plane, $\th^2$ has a kink $\Del \th^2=2\pi k_2/N_2$ on the
$t$-$x$ plane, and $\th^3$ has a kink $\Del \th^3=2\pi k_3/N_3$ on the $x$-$y$
plane.  For such a configuration ($b^I_\mu=0$), the action from the tri-kink
topological term is given by
\begin{align}
 S&=\int \dd x\dd y\dd t\; \frac{\ii}{3} C_{IJK}\varepsilon^{\mu\nu\lambda}
\partial_\mu \theta^I\partial_\nu \theta^J\partial_\lambda \theta^K
\nonumber\\
& =16\pi^3 \ii C_{123} \frac{k_1k_2k_3}{N_1N_2N_3}.
\end{align}
This means that the intersection of the kinks in $\th^1$ and $\th^2$ carries a
$Z_{N_3}$-charge $8\pi^2  C_{123} \frac{k_1k_2}{N_1N_2}$ mod $N_3$.
Since $k_1=0 \sim k_1=N_1$, $C_{123}$ must be quantized:
\begin{align}
 8\pi^2  C_{123} \frac{k_2}{N_2}=0 \text{ mod } N_3,
\ \
 8\pi^2  C_{123} \frac{k_1}{N_1}=0 \text{ mod } N_3. \label{quantization1}
\end{align}
Thus
\begin{align}
  C_{123}=\frac{p_\text{III}}{ (2\pi)^2 2!} \frac{N_1N_2N_3}{N_{123}}, \ \ \ \ p_\text{III}=0,\cdots,N_{123}-1 \label{value1}
\end{align}
where $N_{123}=\text{gcd}(N_1,N_2,N_3)$.  Also we note that $C_{123}$ has only
$N_{123}$ distinct quantized values, \cblue{corresponding} to $N_{123}$ distinct charge
assignments.

Now the physical meaning of the tri-kink topological term is clear: It is well
known that the fluctuations of the kinks will turn a $Z_{N_1}\times
Z_{N_2}\times Z_{N_3}$ symmetry-breaking state into a $Z_{N_1}\times
Z_{N_2}\times Z_{N_3}$ symmetric state with a trivial SPT order.  However, if
we bound a $Z_{N_3}$-charge to the intersection of the kinks in $\th^1$ and
$\th^2$ \etc, the resulting $Z_{N_1}\times Z_{N_2}\times Z_{N_3}$
symmetric state will have a non-trivial SPT order, as we will show below.  In this
way, we can produce $N_{123}$ distinct type-III $Z_{N_1}\times Z_{N_2}\times
Z_{N_3}$ SPT orders, consistent with the group cohomology result.

By integrating out the smooth fluctuations $\theta^I$ and introducing auxiliary gauge fields $a_\lambda^I$ and $\lambda_\mu^I$, we can derive the following bulk dynamical action:
\begin{widetext}
\begin{eqnarray} \label{tri-kink_S}
\mathcal{L}_\text{tri-kink}
&=& {\ii \over2\pi}\varepsilon^{\mu\nu\lambda}\lambda_\mu^I \partial_\nu a_\lambda^I
+ \frac{\ii }{3} C_{IJK}\varepsilon^{\mu\nu\lambda} \big( \lambda_\mu^I \lambda_\nu^J \lambda_\lambda^K %\nonumber\\&&
+ (b_\mu^I-\lambda_\mu^I) (b_\nu^J-\lambda_\nu^J) (b_\lambda^K-\lambda_\lambda^K) \big)
+\frac{1}{2}(b_\mu^I-\lambda_\mu^I)^2+\mathcal{L}_\text{Maxwell}^b.\;\;\;\;\;\;\;
\end{eqnarray}

\cblue{The derivation from Eq.(\ref{tri-kink}) to Eq.(\ref{tri-kink_S}) is preserved in Appendix \ref{derivation} with details.}
Interestingly, the field strength of gauge field $a_\mu^I$ is formally akin to a non-Abelian gauge field and its \cblue{infinitesimal} gauge transformation should be modified as:
\begin{eqnarray}
a_\mu^I \rightarrow a_\mu^I +\partial_\mu f^I-4\pi C_{IJK}\left(g^J \lambda_\mu^K+ \frac{1}{2}g^J\partial_\mu g^K \right) ;
\quad b_\mu^I \rightarrow b_\mu^I +\partial_\mu g^I ;
\quad \lambda_\mu^I \rightarrow \lambda_\mu^I +\partial_\mu g^I.
\end{eqnarray}

\end{widetext}

% and bulk response theory
\subsection{Saddle point approximation and internal gauge theory} \label{sec:saddle-approx}

If we assume the field $b_\mu^I$ has a weak fluctuation, we can apply the saddle point approximation for $b_\mu^I$. The saddle point equation reads:
\begin{align}
& C_{IJK}\varepsilon^{\mu\nu\lambda}   (b_\nu^J-\lambda_\nu^J) (b_\lambda^K-\lambda_\lambda^K)
+ \cblue{(b_\mu^I - \lambda_\mu^I)}+
\nonumber\\
&\ \ \ \ \ \ \ \ \text{higher order terms}
=0,
\end{align}
clearly $b_\mu^I=\lambda_\mu^I$ is a stable saddle point.
\cblue{Since the $\lambda$ field is a Lagrangian multiplier and $b$ is a more-restricted $U(1)$ field, %compact
we should replace $\lambda$ by $b$.}
At the level of this approximation, we can simplify the bulk effective action by:
\begin{align} \label{effective} %\label{eq:2+1DSPT}
\mathcal{L}_\text{eff}=\frac{\ii \varepsilon^{\mu\nu\lambda}}{2\pi} b_\mu^I \partial_\nu a_\lambda^I + \frac{\ii C_{IJK}}{3} \varepsilon^{\mu\nu\lambda} b_\mu^I b_\nu^J b_\lambda^K , %+ \dots +\mathcal{L}_\text{Maxwell}^\lambda,
\end{align}
and
with the gauge redundancy given by:
\begin{align}
& b_\mu^I \rightarrow b_\mu^I +\partial_\mu g^I ; \nonumber\\
& a_\mu^I \rightarrow a_\mu^I +\partial_\mu f^I-4\pi C_{IJK} \left( g^J b_\mu^K + \frac{1}{2}g^J\partial_\mu g^K\right) . \label{gauge1}
\end{align}
%Here the gauge transformation reveals that $\ee^{\ii f^I}$, $g^I$, and $b^I$ are globally defined. However,
%${f^I}$ needs not to be globally defined.
%Since $\dd^2 f$ is associated to the local $U(1)$ charge,
%the non-global defined ${f^I}$ allows the
%2D surface integral of $\Ointint \dd a/(2\pi)$ to be quantized as any integer as the number of the net $U(1)$ charge.
%On the other hand, $b^I$ are globally defined, thus $\Ointint \dd b/(2\pi)=0$. To summarize,
We also have the global constraints:
\begin{align}
\Ointint \dd a/(2\pi)  \in \mathbb{Z}, \ \ \ \
\Ointint \dd b_I/(2\pi) \in \mathbb{Z}
\end{align}
%In Sec.\ref{sec:partition-GSD}, we will use a more intuitive physical picture from
%the $U(1)$ charge and the vortex number to derive the same constraints.

Similar to the $1+1$D case, there is a rigorous way to compute the quantization of coefficients $C_{123}$ protected by global $Z_{N_1}\times Z_{N_2}\times Z_{N_3}$ symmetry. Let us add a coupling term to the external gauge field $A^I$.

\begin{equation}
\mathcal{L}_\text{coupling}=\cblue{\ii}  {A}^I_{\mu} j^{\mu}_I=\frac{\ii }{2\pi}\varepsilon^{\mu\nu\lambda} A^I_\mu\partial_\nu a^I_\lambda ,
\end{equation}
Again, $A^I$ are $Z_{N_I}$ symmetry twists, thus $A^I$ must be a flat connection with $\dd A^I =0$ and $\oint A^I=2\pi n_I/N_I$. Similar to the $1+1$D case, since $\int \dd x \dd y \dd t \mathcal{L}_\text{coupling}$ must be invariant under gauge transformation Eq.(\ref{gauge1}), $C_{123}$ can not take arbitrary value, and a short calculation gives rise to exactly the same condition Eq.(\ref{quantization1}).

It turns out that the above gauge transformation corresponds to a \emph{non-semisimple Lie algebra} of symmetry.
We will discuss a generic class of such Lie algebra, called \emph{the symmetric-self dual Lie algebra} in Appendix \ref{non-semi-Simple}.
%\textbf{Move the following non-semisimple Lie algebra to appendix or a later section. We try to avoid complicated thing at this stage. -- XGW}
\begin{comment}
\cblue{If we write the gauge connection in terms of its gauge field components and its generators:
\begin{eqnarray}
&&\tilde{a}_\mu^\alpha T^\alpha \equiv b_\mu^I  X_I + a_\mu^I H_I^*.\\
&& (\tilde{a}_\mu^1 T^1,\tilde{a}_\mu^2 T^2, \tilde{a}_\mu^3 T^3)=
(b_\mu^1  X_1,b_\mu^2  X_2,b_\mu^3 X_3),\\
&& (\tilde{a}_\mu^4 T^4,\tilde{a}_\mu^5 T^5, \tilde{a}_\mu^6 T^6)=
(a_\mu^1  H_1^*,a_\mu^2  H_2^*, a_\mu^3 H_3^*).
\end{eqnarray}
Here $\alpha=1,\dots,6$ and $I=1,\dots,3$.}

The corresponding generators $H^I$ and $X_I$
satisfy:
\begin{align} \label{eq:Lie_algebra}
[H_I^*,H_J^*]=[H_I^*,X_J]=0; \quad [X_I,X_J]=C_{IJK}H_K^*,
\end{align}
where $C_{IJK}$ serves as the structure constant now.
\cblue{The full Lie algebra consists of an Abelian Lie algebra $\mathcal{X}(X)$ with a central %Abelian
extension by another Abelian Lie algebra $\mathcal{H}^*(H^*)$.}
%
%Here $\mathcal{X}(X)$ contains the set of generators $X^I$, and $\mathcal{H}(H)$ contains the set of generators $H^I$.
\end{comment}
%
In Sec.\ref{sec:partition-GSD},
we will define a rigorous SPT internal gauge theory path integral, and
we confirm that the GSD of our theory is unique on a closed manifold, GSD=1, just like the SPT state.
We will also derive the \emph{SPT invariant} by coupling the internal gauge theory to semi-classical probed field $A$ claimed in Ref.\cite{JuvenSPT1}, which suggests that Eq.(\ref{value1}) indeed gives rise to $N_{123}$ distinguishable SPT phases.
%

%We will leave the more generic
%properties of the non-semisimple Lie algebra into Appendix \ref{non-semi-Simple}. Our immediate goal is to confirm the bulk probed field theory
%as the desired SPT invariant in Ref.\cite{JuvenSPT1}.

\section{A 3+1D generalization}
The above tri-kink topological term can be generalized into higher dimensions as well, such as
a quad-kink topological action in 3+1D:
\begin{eqnarray} \label{q-kink}
\mathcal{L}_\text{q-kink}=&&{1\over2}(\partial_\mu \theta^I_\s + b_\mu^I)^2
+
\frac{\ii }{4} C_{IJKL}\varepsilon^{\mu\nu\lambda \sigma}(\partial_\mu \theta^I_\s + b_\mu^I) \nonumber\\
&&\cdot(\partial_\nu \theta^J_\s + b_\nu^J)(\partial_\lambda \theta^K_\s + b_\lambda^K)(\partial_\sigma \theta^L_\s + b_\sigma^L).
\end{eqnarray}
The quantization condition on $C_{IJKL}$ can be worked out in a similar way, and finally we obtain
$ C_{1234}=\frac{ p_\text{IV} }{ (2\pi)^3 3!} \frac{N_1N_2N_3 N_4}{N_{1234}}$,
where $p_\text{IV}$ is an integer on $p_\text{IV}=0,\cdots,N_{1234}-1$.

For example, in $3+1$D, we can use the following
quartic-kink term to describe the so-called type-IV SPT state. Parallel to our previous derivation in Sec.\ref{sec:Tri-kink topological}, we can derive the
SPT bulk dynamical action:

\begin{widetext}
%\begin{eqnarray}
\begin{equation}
\label{q-kink_S}
\mathcal{L}_\text{q-kink}
= {\ii \over2\pi}\varepsilon^{\mu\nu\lambda \rho}\lambda_\mu^I \partial_\nu a_{\lambda \rho}^I
- \frac{\ii }{4} C_{IJKL}\varepsilon^{\mu\nu\lambda \rho}
\big(\lambda_\mu^I \lambda_\nu^J \lambda_\lambda^K \lambda_\rho^L %\nonumber\\&&
- (b_\mu^I-\lambda_\mu^I) (b_\nu^J-\lambda_\nu^J)(b_\lambda^K-\lambda_\lambda^K)(b_\rho^L-\lambda_\rho^L) \big)
+\frac{1}{2}(b_\mu^I-\lambda_\mu^I)^2 + \mathcal{L}_\text{Maxwell}^b %+ \dots. \;\;%
\end{equation}
%\end{eqnarray}
and its gauge transformation:
\begin{eqnarray}
 a_{\mu\nu}^I &\rightarrow a_{\mu\nu}^I +\partial_\mu f^I_\nu-\partial_\nu f^I_\mu \cblue{+24} \pi C_{IJKL} g^J \lambda_\mu^K \lambda_\nu^L  \cblue{+\dots};
\quad b_\mu^I \rightarrow b_\mu^I +\partial_\mu g^I + \dots;
\quad \lambda_\mu^I \rightarrow \lambda_\mu^I +\partial_\mu g^I + \dots.
\end{eqnarray}
\end{widetext}
Here $\mathcal{L}_\text{Maxwell}^b$ terms contain non-topological Maxwell term.
%If we perform the saddle-point approximation, we obtain $\lambda_\mu^I=b_\mu^I$, and we replace the Lagrange multiplier $\lambda_\mu^I$ to a more restricted $b_\mu^I$.
If we further apply the saddle-point approximation, we obtain:
\begin{align} \label{effective3D}
\mathcal{L}_\text{eff}=\frac{\ii \varepsilon^{\mu\nu\rho\sigma}}{4\pi} b_\mu^I \partial_\nu a_{\sigma\rho}^I \cblue{-} \frac{\ii C_{IJKL}}{4} \varepsilon^{\mu\nu\sigma\rho} b_\mu^I b_\nu^J b_\sigma^K b_\rho^L. %+ \dots
%+\mathcal{L}_\text{Maxwell}^\lambda.
\end{align}
The corresponding infinitesimal gauge transformation(we only keep the leading order term here and use $\dots$ to represent higher order terms) of arbitrary functions $f$ and $g$ reads:
\begin{align}
 a_{\mu\nu}^I &\rightarrow a_{\mu\nu}^I +\partial_\mu f^I_\nu-\partial_\nu f^I_\mu \cblue{+24} \pi C_{IJKL} g^J b_\mu^K b_\nu^L  \cblue{+\dots}, \nonumber\\
b_\mu^I &\rightarrow b_\mu^I +\partial_\mu g^I \cblue{+\dots}.
\end{align}
Here $g^I$ and $b^I$ are globally defined,
but $f^I$ is not globally defined.
The analogous global constraint can be derived:
\begin{align}
\Ointintint \dd a/(2\pi)  \in \mathbb{Z},\ \ \ \ \
\Ointint \dd b_I/(2\pi)  \in \mathbb{Z}.
\end{align}
%see also Sec.\ref{sec:partition-GSD}.

%Here $\dots$ implies beyond the saddle-point approximation and the unspecified global constraints.

% bulk response theory
\section{Partition function, GSD and SPT invariants computed from the SPT internal gauge theory} \label{sec:partition-GSD}

\color{black}

Here we will analytically show the path integral definition of
internal gauge field theory, Eq.(\ref{eq:1+1DSPT}) for 1+1D,
Eq.(\ref{effective}) for 2+1D,
Eq.(\ref{effective3D}) for 3+1D.
In particular, we will show three key issues:\\
$\bullet (i)$ Define the partition function $\mathbf{Z}$ using field theory path integral.\\
$\bullet (ii)$ Derive the SPT invariants of semi-classical flat probed field theory in Ref.\cite{JuvenSPT1}
by coupling the SPT internal gauge theory to probed fields $A$.\\
$\bullet (iii)$  The internal gauge theory on any compact closed spatial manifold
has a unique ground state, namely GSD=1. This means that
the absolute value of the phase space volume ratio between the case with topological term
and the case without topological term: $|\frac{\mathbf{Z}}{\mathbf{Z}(p=0)}|=1$.

%We will focus on the 2+1D example Eq.(\ref{effective}).
This procedure also applies to the SPT internal gauge field theory in any other dimensions.
We know that the SPT state has no intrinsic topological order and the SPT's GSD=1 on any compact closed spatial manifold.
Therefore, this GSD computation serves as the consistency check that the internal field theory shows a gapped phase with nontrivial
symmetry transformation --- the internal gauge field theory realizes SPT state.

We emphasize that knowing the field theory action is \emph{not enough} to fully understand the SPT field theory.
We stress that defining the partition function $\mathbf{Z}$ using field theory path integral
is \emph{necessary} to fully understand the SPT field theory. Below we especially remark the global constraints of fields in order
to define the SPT path integral.
%\begin{eqnarray}
The partition function in terms of the path integral form with a total spacetime dimension $d$ is
\begin{widetext}
\begin{equation} \label{eq:bda+bbb_SPT}
\mathbf{Z}=
\int [Db] [Da]
\exp\big(\int  \big( \frac{\ii}{2\pi} b^I \wedge \dd a^I
+\frac{\ii (-1)^{d-1} C_{IJK \dots}}{d}  b^I \wedge b^J  \wedge b^K \wedge \dots \big) \big),
\end{equation}
\end{widetext}
here $I,J,K, \dots \in \{ 1,2,3, \dots, d \}$.
Here $b$ is 1-form, $a$ is $(d-2)$-form, and $ f= \dd a$ is $(d-1)$-form.
In the presence of symmetry-twist semi-classical background 1-form gauge field $A$, we can write
the partition function $\mathbf{Z}$ as
\begin{widetext}
\begin{eqnarray}
&&\mathbf{Z}=\int [Db] [Da]
\exp\big(\int  \big( \frac{\ii}{2\pi} (b^I-A^I) \wedge \dd a^I
+\frac{\ii (-1)^{d-1} C_{IJK \dots}}{d}  b^I \wedge b^J  \wedge b^K \wedge \dots \big) \big) \nonumber\\
&&
\label{eq:bAf+bbb_SPT}
=\int [Db] [Df]
\exp\big(\int  \big( \frac{\ii}{2\pi} (b^I-A^I) \wedge f^I
+\frac{\ii (-1)^{d-1} C_{IJK \dots}}{d}  b^I \wedge b^J  \wedge b^K \wedge \dots \big) \big),
\end{eqnarray}
\end{widetext}
with the field strength of charges $f \equiv \dd a$.
Importantly, we view $b^I$ and $a^I$ all \emph{dynamical internal gauge fields}, so they are involved in the path integral measure.\\

Now let us define this path integral properly.
Let us impose the constraints for this field function in the path integral, based on the dual equivalent theory using the non-linear $\sigma$-model. %Recall that
%the $b$ is related to the vortex charge and current via $* j_{\text{vortex}} \equiv (\dd^2 \theta_{\text{v}})/(2\pi)=\dd b/(2\pi)$.
%On a compact closed 2-surface, the total net number of vortices is zero (namely, vortices are cancelled with anti-vortices):
%\begin{eqnarray} \label{eq:global_b}
%&& \Ointint *j_{\text{vortex}} =\Ointint \dd b/(2\pi)  =0.
%\end{eqnarray}
%On the other hand,
We recall that the $a$ is related to the \emph{current density} $j$ specified by the $U(1)$ or $Z_N$ charge, where
we have the total number of charges quantized:
\begin{eqnarray}  \label{eq:global_a}
 \Ointint *j =\Ointint \dd a/(2\pi) =\Ointint f /(2\pi)  \in \mathbb{Z} ,
\end{eqnarray}
The current density $*j $ is a $(d-1)$-form,
thus $\Ointint$ of $\dd a$ represents the surface integral of a $(d-1)$-closed manifold, such as a 1-surface for 1+1D spacetime, 2-surface for 2+1D spacetime.

Now we integrate over the field variable $f$ for the partition function Eq.(\ref{eq:bAf+bbb_SPT}),
which procedure analogous to the discrete Fourier summation yields a constraint:
\begin{eqnarray}
\sum_{n \in \mathbb{Z} } e^{\ii \varphi n} =\delta(\varphi \text{ mod } {2\pi}).
\end{eqnarray}
For $\int [Df]
e^{\int   \frac{\ii}{2\pi} (b^I-A^I) \wedge f^I }$ with $\Ointint f /(2\pi)  \in \mathbb{Z} \text{ or } \mathbb{Z}_N$, we obtain an analogous constraint  on a 1D loop:
\begin{eqnarray}
&&\oint (b^I-A^I)=0 \text{ mod } {2\pi} \\
%\end{eqnarray}
%\begin{eqnarray}
&&\Rightarrow \oint b^I= \oint A^I \text{ mod } {2\pi}= \frac{2\pi n_I}{N_I} \text{ mod } {2\pi}. \label{eq:global2}
\end{eqnarray}
The first line constraint is true for both $U(1)$ charge and $Z_N$ charge.
The second line constraint Eq.(\ref{eq:global2}) with $n_I \in \mathbb{Z}$ is an additional constrain if $\Ointint f /(2\pi)  \in \mathbb{Z}_{N_I}$ for our case of
discrete $Z_N$ charge for SPT state with $Z_N$-symmetry.
We can still view $b$-field sa a $U(1)$ connection but with a constraint from the $Z_N$ symmetry-twist probed-field $A$.
This means that the internal gauge field $b$ is subject to the global constraint from the semi-classical symmetry-twist probed field $A$.
%Thus far
After integrating out the $f$, the partition function Eq.(\ref{eq:bAf+bbb_SPT}) subject to the global constraint Eq.(\ref{eq:global2})
of the symmetry-twist fields $A$ becomes
\begin{widetext}
\begin{eqnarray} \label{eq:ZSPTinv}
%\begin{equation}
\mathbf{Z}
= \int [Db]
\exp\big(\int
\frac{\ii (-1)^{d-1} C_{I_1 I_2 \dots I_d}}{d}
b^{I_1} \wedge b^{I_2}  \wedge  \dots \wedge b^{I_d}\big)
=\exp\big(\int  \big( \frac{\ii (-1)^{d-1}  C_{I_1 I_2 \dots I_d}}{d}  A^{I_1} \wedge A^{I_2}  \wedge  \dots \wedge A^{I_d} \big) \big).\;\;\;\;\;
%\end{equation}
\end{eqnarray}
Thus so far by using SPT internal gauge theory path integral, we have recovered the SPT invariant of Ref.\cite{JuvenSPT1} claimed in the item $(ii)$.
\end{widetext}
Next, without losing generality, let us take $2+1$D SPT as an example, with an explicit $C_{IJK}=\frac{1}{(2 \pi)^2 2!} \frac{N_1 N_2 N_3\;
p_{\text{III} }}{N_{123}}$.
Let us do the explicit partition function calculation on the two topologies, a sphere and a torus respectively, by comparing
 the nontrivial class ${\mathbf{Z}}$ to the trivial class ${\mathbf{Z}(p_{\text{III}}=0)}$.
For each calculation below we will fix a particular set of $n_I$ for the global constraint Eq.(\ref{eq:global2}). %boundary condition

{\bf The 1st topology}: On a spatial sphere $S^2$ with a time loop $S^1$, there is only a non-contractible loop along the time direction.
So there is only a nonzero $n_I$ for the global constraints in Eq.(\ref{eq:global2}), %boundary conditions
and other $n_J$ must be zeros. We have:
\begin{align}
\frac{\mathbf{Z}}{\mathbf{Z}(p_{\text{III}}=0)}
&=\frac{\exp\big(  \ii\frac{2 \pi  p_\text{III} }{N_{123}}  0 \cdot 0 \cdot n_I   \big)}{1}=1.
\end{align}

{\bf The 2nd topology}: On a spacetime $T^3$ torus, without losing generality, let us assume, $A^1$, $A^2$, $A^3$ along $x,y,t$-directions
have nontrivial global constraints with some generic $n_1$, $n_2$ and $n_3$.
For example, analogous to Sec.\ref{sec:Tri-kink topological}'s setup, we can assume $\dd x^\mu=\dd x$, $\dd x^\nu=\dd y$, $\dd x^\rho=\dd t$.
\begin{equation}
\frac{\mathbf{Z}}{\mathbf{Z}(p_{\text{III}}=0)}=\frac{\exp\big(   \ii p_\text{III}  \frac{2\pi \cdot n_1n_2 n_3}{ N_{123}} \big)}{1}
=\ee^{\ii p_\text{III}  \frac{2\pi \cdot n_1n_2 n_3}{ N_{123}}}.
\end{equation}

Since for both on a sphere and on a torus, the absolute value of the above, $|\frac{\mathbf{Z}}{\mathbf{Z}(p_{\text{III}}=0)}|$, measures the GSD ratio between the nontrivial phases and the trivial insulator.
Since the trivial insulator has GSD=1 here, all other phases have GSD=1, so the $p_{\text{III}} \neq 0$ phase is a generic SPT state.

We thus confirm that the path integral Eq.(\ref{eq:bda+bbb_SPT}) with dynamical variables describes nontrivial type III SPT states in 2+1D.
The same procedure can be generalized to other dimensions, such as
Eq.(\ref{eq:1+1DSPT}) as SPT states in 1+1D and Eq.(\ref{effective3D}) as SPT states in 3+1D.
The GSD for these theories defined by the partition function is 1.
The procedure works in more general closed topology, we thus show the claim in the item $(iii)$.
\color{black}

\cblue{One further extension of our work is to study the \emph{duality}\cite{duality}
between SPT (which is non-topologically ordered) and dynamical topological gauge theory (which is topologically ordered).
More precisely,
we can start from the SPT internal gauge theory path integral of Eq.(\ref{eq:bAf+bbb_SPT})
and then \emph{dynamically gauge} the theory to a dynamical topological gauge theory equivalent to the Dijkgraaf-Witten theory \cite{Dijkgraaf:1989pz}.
In Appendix.\ref{dynamical:DW}, we will outline such a procedure using field theory path integral, and we will
propose the \emph{continuous dynamical topological gauge theory} dual to the \emph{Dijkgraaf-Witten theory with a discrete gauge group}.
}

\section{Edge theory} %velocity matrix
The bulk effective field theory can also describe interesting edge physics. For the 1+1D case, by integrating out the \cblue{Lagrange multiplier}
fields $a^I$ in Eq.(\ref{eq:1+1DSPT}),
the corresponding edge theory takes a very simple form:
\begin{eqnarray}
\mathcal{L}_\text{edge}^0=\frac{\ii }{2} C_{IJ} \varphi^I \partial_0 \varphi^J,
\end{eqnarray}
with scalar fields $\varphi^I$ define the gauge transformation $\varphi^I \to \varphi^I - g^I$ to cancel the
gauge transformation of $b^I \to b^I+ \dd g^I$.
which is nothing but a quantized topological term for a quantum \cblue{mechanical} system with degenerate ground states.
%\cblue{{\bf Juven: comment about the Euclidean/Minkowski? $\partial_0$ for the Euclidean time?}}
Such a Berry phase implies the following quantization condition:
\begin{eqnarray}
[\varphi^1, \varphi^2]=\frac{\ii}{C_{12}}=\frac{2\pi \ii N_{12}}{p_{\text{II}}N_1N_2}=\frac{2\pi \ii  }{ {p_{\text{II}}} N^{12}},
\end{eqnarray}
\cblue{Here $N^{12}$ is defined as the least common multiplier ($\text{lcm}$) where $N^{12} \equiv \text{lcm}(N_1,N_2)={N_1N_2}/N_{12}$.}
%%%%%%%%%%%%%
%Since both $\varphi^1$ and $\varphi^2$ are compact scalar fields, we should consider the quantization condition between \cblue{$W_{\varphi^1}=e^{\ii N_1\varphi^1}$}
%and \cblue{$W_{\varphi^2}=e^{\ii N_2\varphi^2}$}. It is easy to find:
%\cblue{
%\begin{eqnarray}
%W_{\varphi^1}W_{\varphi^2}=e^{-\frac{2\pi \ii N_{12}}{p_\text{II}}}W_{\varphi^2}W_{\varphi^1}, \; I=1,2
%\end{eqnarray}
%}
%which is the $Z_{\frac{N_1N_2}{p_\text{II}}}$ Heisenberg algebra and requires a $\frac{N_1N_2}{p_\text{II}}$ dimensional representation.
%%%%%%%%%%%%%%
%Here we suppose
Due to the compactification and the quantization constraint, shown in Appendix \ref{sec:GSD2}, %for verification
the symmetry generators are \cblue{$S_{\varphi^1}=e^{\ii N_1\varphi^1 \frac{p_{\text{II}}}{N_{12}}}$}
and \cblue{$S_{\varphi^2}=e^{\ii N_2\varphi^2  \frac{p_{\text{II}}}{N_{12}} }$}.
%\begin{comment}
It is straightforward to check that
$
S_{\varphi^I} (\int \dd t \mathcal{L}_\text{edge}^0)  S_{\varphi^I}^{-1}=(\int \dd t \mathcal{L}_\text{edge}^0) + 2 \pi \cdot \text{integer},
$
so the partition function $\mathbf{Z}=\int D {\varphi^1} D {\varphi^2} e^{-\int \dd t \mathcal{L}_\text{edge}^0} \ $is invariant under the symmetry transformation $S_{\varphi^I}$.
%\end{comment}
We find that the symmetry is realized in a projective representation manner on the 0D edge, because the symmetry generators do not commute:
\cblue{
\begin{eqnarray}
S_{\varphi^1} S_{\varphi^2}=e^{-\frac{2\pi \ii p_\text{II}}{ N_{12}}} S_{\varphi^2} S_{\varphi^1}.
\end{eqnarray}
}
Here $p_\text{II}$ is defined as a  $p_\text{II}$ (mod ${N_{12}}$) variable.
 If $\gcd(p_\text{II}, { N_{12}})=1$, it is the $Z_{N_{12}}$ Heisenberg algebra and requires a $N_{12}$-dimensional representation
for the symmetry generators $S_{\varphi^1}$ and $S_{\varphi^2}$.
This implies the 0+1D edge mode of the ground state has a ${N_{12}}$-fold degeneracy, consistent with the edge mode physics analysis via the dimensional reduction approach in Ref.\cite{Wang:2014tia}.
In general, even if $\gcd(p_\text{II}, { N_{12}}) \neq 1$, we have a generic $\frac{N_{12}}{\gcd(p_{\text{II}},{N_{12}})}$-dimensional representation for
the symmetry generators, thus the zero mode degeneracy is
\begin{eqnarray}
\GSD=\frac{N_{12}}{\gcd(p_{\text{II}},{N_{12}})}.
\end{eqnarray}

\cblue{For the 2+1D bulk system with its 1+1D edge theory, we have an analogous derivation as follows.}
Integrating out $a_\mu$ leads to the constraint:
\begin{equation}
\varepsilon^{\mu\nu\lambda}\partial_\mu \lambda_\nu^I=0,
\end{equation}
The constraint can be solved by requiring:
\begin{equation}
\lambda_\nu^I=\partial_\nu \varphi^I,
\end{equation}
We see that $\mathcal{L}_\text{eff}$ is nothing but a total derivative:
\begin{eqnarray}
\mathcal{L}_\text{eff}=\frac{\ii }{3} C_{IJK}\varepsilon^{\mu\nu\lambda} \partial_\mu \varphi^I \partial_\nu \varphi^J \partial_\lambda \varphi^K,\label{boundary}
\end{eqnarray}
which actually describes a $1+1$D edge with effective action:
\begin{eqnarray}
\mathcal{L}_\text{edge}^1=\frac{\ii }{3} C_{IJK}\varepsilon^{\mu\nu} \varphi^I \partial_\mu \varphi^J \partial_\nu \varphi^K.\label{edge}
\end{eqnarray}

The higher dimensional generalization is also straightforward, e.g., the type-IV SPT in $3+1$D can have a $2+1$D edge theory described by:
\begin{eqnarray}
\mathcal{L}_\text{edge}^2=\frac{\ii }{4} C_{IJKL}\varepsilon^{\mu\nu\rho} \varphi^I \partial_\mu \varphi^J \partial_\nu \varphi^K \partial_\rho \varphi^L.\label{edge3D}
\end{eqnarray}
The gapless nature of these boundary terms can be proved via dimension reduction to the 1+1D case we discussed at the beginning of this section. Finally, we note that if we view $\varphi^I$ as scaling dimension zero fields, $\mathcal{L}_\text{edge}^1$ and $\mathcal{L}_\text{edge}^2$ can be regarded as a fractionalized version of $O(3)$ and $O(4)$ topological theta terms. For future work, it would be of great interest to understand the underlying conformal field theory described by these fractionalized theta terms.

\section{Topological field theory for Dijkgraaf-Witten lattice model}
%(\ie the cocycle-twisted gauge theory)
%Topological gauge theory: Dynamically gauging SPT internal field theory dual to Dijkgraaf-Witten theory
\label{dynamical:DW}

In Sec.\ref{sec:partition-GSD}, we had established the SPT field theory by defining the SPT path integral.
It is known that there exists a \emph{duality}\cite{duality}
between SPT (which is non-topologically ordered) and dynamical topological gauge theory (which is topologically ordered).
More precisely,
we can start from the SPT internal gauge theory path integral of Eq.(\ref{eq:bAf+bbb_SPT})
and then \emph{dynamically gauge} the theory by coupling the SPT matter field to external probed fields $A$, and
make the $A$ dynamical gauge fields. This procedure of gauging SPT with a finite symmetry group
in principle yields a dynamical topological gauge theory equivalent to the Dijkgraaf-Witten theory\cite{Dijkgraaf:1989pz}.
Here we describe such a procedure using field theory path integral, and we
propose some \emph{continuous dynamical topological gauge theory} dual to the \emph{Dijkgraaf-Witten theory with a discrete gauge group}.

Naively, one approach is starting from the path integral Eq.(\ref{eq:bAf+bbb_SPT}), if we promote the semi-classical probed field $A$ to a dynamical field by including the
path integral measure $[DA]$, we obtain:
\begin{widetext}
\begin{eqnarray} \label{eq:DW_gauge1}
&&\mathbf{Z}=\int [Db] [Da] [DA]
\exp\big(\int  \big( \frac{\ii}{2\pi} (b^I-A^I) \wedge \dd a^I + \frac{\ii C_{IJK \dots}}{N} b^I \wedge b^J  \wedge b^K \wedge \dots \big) \big)
\end{eqnarray}
\end{widetext}

One can see that if $A$ is still subject to some global constraint:
\begin{eqnarray}
 \oint A^I \text{ mod } {2\pi}= \frac{2\pi n_I}{N_I} \text{ mod } {2\pi}.
\end{eqnarray}
but now $n_I \in \mathbb{Z}_{N_I}$ needs \emph{not} to be fixed. The dynamical gauge theory of $A$ would sum
over all possible $n_I$.
If we compute the GSD of this field theory on a spacetime manifold, then we essentially reproduce the same calculation
using the group cohomology cocycle while summing over all possible group elements $n_I \in \mathbb{Z}_{N_I}$.
Eq.(\ref{eq:DW_gauge1}) can produce the same physical observables such as GSD of Dijkgraaf-Witten theory.
This suggests that Eq.(\ref{eq:DW_gauge1}) can be an equivalent description of Dijkgraaf-Witten theory.

Another approach to obtain the dynamical gauge theory is through the minimal coupling the internal gauge field $a$ to the external gauge field $A$,
and then integrating out all the internal gauge fields $a$ and $b$. We describe it below.

{\bf 2+1D}: Now, we are ready to discuss the bulk response theory. The external probe gauge field $A_\mu^I$ will couple to the internal charge current in a standard way:
\begin{equation}
\mathcal{L}_\text{coupling}=\cblue{\ii}  {A}^I_{\mu} j^{\mu}_I=\frac{\ii }{2\pi}\varepsilon^{\mu\nu\lambda} A^I_\mu\partial_\nu a^I_\lambda ,
\end{equation}
However, since ${A}^I_{\mu}$ is in the Higgs phase with $Z_{N_I}$ charge condensation, we need to introduce a BF term\cite{SPTCS3} for response guage field ${A}^I_{\mu}$ as well:
\begin{equation}
\frac{\ii N_I}{2\pi}\varepsilon^{\mu\nu\lambda}B_\mu^I\partial_\mu A_\nu^I,
\end{equation}
Actually such a term is crucial for maintaining the gauge invariance for the total action.(It is easy to check that $\mathcal{L}_\text{coupling}$ is not gauge invariant under the gauge transformation of $a^I_\mu$ and we need to shift $B_\mu^I$ to restore the gauge invariance.)
%This can be imposed by adding a Lagrangian multiply term in the action:
%\begin{equation}
% \mathcal{L}_\text{constraint}=\frac{\ii }{2\pi}B_\lambda^I\varepsilon^{\mu\nu\lambda}\partial_\mu A_\nu^I,
%\end{equation}
Finally, by integrating out the internal gauge field $a_\mu^I$ and \cblue{$b_\mu^I$}, we end up with an effective action
$\frac{\ii N_I}{2\pi}\varepsilon^{\mu\nu\lambda}B_\mu^I\partial_\mu A_\nu^I+\frac{\ii }{3} C_{IJK}\varepsilon^{\mu\nu\lambda} A_\mu^I A_\nu^J A_\lambda^K$:
\begin{align}
\mathcal{L}_\text{response} &=
\frac{\ii N_I}{2\pi}\varepsilon^{\mu\nu\lambda}B_\mu^I\partial_\mu A_\nu^I+\frac{\ii p_\text{III}N_1N_2N_3}{\cblue{(2\pi)^2} N_{123}} \varepsilon^{\mu\nu\lambda} A_\mu^1 A_\nu^2 A_\lambda^3.
\end{align}
%1/ {(2\pi)^2}= 3!/{(24\pi)}
%%%%%
If we view $A_\mu^I$ as background gauge fields describing the symmetry twists on the boundary, the above action is equivalent to the SPT invariants \Eq{SPTinv2DF}. However, if we view both $A_\mu^I$ and $B_\mu^I$ as dynamical gauge fields, the above action \cblue{potentially}
describes non-Abelian Berry phases, though the original global symmetry is Abelian and all the gauge fields are Abelian in its own sectors.
The whole Lie algebra becomes non-Abelian feature due to the central extension Eq.(\ref{eq:Lie_algebra}).
\cblue{It will be interesting to verify whether the fully-dynamical topological gauge theory
is equivalent to the Dijkgraaf-Witten gauge theory\cite{Dijkgraaf:1989pz}.
Our word of caution is that the non-semi-simple Lie algebra detailed in Appendix \ref{non-semi-Simple}
suggests %a future investigation is necessary.
a more conservative %and negative
side of this claim.
It is also likely that method beyond the-saddle-point approximation is required to capture the global constraints and missing pieces that we may omit in
Eqs.(\ref{tri-kink_S}) and (\ref{effective}).
}

%{\bf [Explain more] A detailed calculation for the ground state degeneracy and the braiding T and S matrices will be presented elsewhere. }

%To this end,
%\cblue{In summary}, we discuss the physical mechanism for type III Abelian SPT states in 2+1D and propose a bulk dynamical effective action beyond the
%\cblue{well-known} Chern-Simons theory. We further show that such an action indeed produce the correct SPT invariants for type III Abelian SPT states in 2+1D.

%\subsection{3+1D}

{\bf 3+1D}: Similarly, we can discuss the bulk response theory. The external probe gauge field $A_\mu^I$ will couple to the internal charge current in a standard way:
\begin{equation}
  \mathcal{L}_\text{coupling}=\cblue{\ii} {A}^I_{\mu} j^{\mu}_I=\frac{\ii }{4\pi}\varepsilon^{\mu\nu\rho\sigma} A^I_\mu\partial_\nu a^I_{\rho\sigma} ,
\end{equation}
Similar to the 2+1D case, we also need to introduce \cblue{a BF term} to describe the $Z_{N_I}$ external gauge field in \cblue{3+1D}:
\begin{equation}
\frac{\ii N_I}{4\pi}\varepsilon^{\mu\nu\rho\sigma}B_{\mu\nu}^I\partial_\rho A_\sigma^I,
\end{equation}
By integrating out the internal gauge field $a_{\mu\nu}^I$ and $\lambda_\mu^I$, we end up with an effective action:
\begin{align} \label{eq:3+1dAAAA}
\mathcal{L}_\text{response}&=\frac{\ii N_I}{4\pi}\varepsilon^{\mu\nu\rho\sigma}B_{\mu\nu}^I\partial_\rho A_\sigma^I\nonumber\\&\cblue{-}
\frac{\ii }{4} C_{IJKL}\varepsilon^{\mu\nu\rho\sigma} A_\mu^I A_\nu^J A_\rho^K A_\sigma^L + \dots.
\end{align}
%\begin{align}
%\mathcal{L}_\text{response} &=\frac{\ii N_I}{4\pi}\varepsilon^{\mu\nu\rho\sigma}B_{\mu\nu}^I\partial_\rho A_\sigma^I \nonumber\\& \cblue{-}
%\frac{\ii p_\text{IV}N_1N_2N_3N_4}{ \cblue{(2 \pi)^3} N_{1234}} \varepsilon^{\mu\nu\rho\sigma} A_\mu^1 A_\nu^2 A_\rho^3 A_\sigma^4 + \dots ,
%\end{align}
% (2 \pi)^3=8\pi^3
\cblue{We warn the reader that there is a potential danger to view Eq.(\ref{eq:3+1dAAAA}) as
the dynamical topological gauge theory, as one needs to further confirm the physical properties such as topological GSD and braiding statistics
must match with the 3+1D Dijkgraaf-Witten topological gauge theory\cite{Dijkgraaf:1989pz} computed in Ref.\cite{Wang:2014oya}.
We will leave the study of topological gauge theories for future work.
The minimum claim of our approach is that viewing the $B$ field as a Lagrangian multiplier constrains the flatness of $A$ with $\dd A =0$,
we essentially derive the SPT invariant in terms of the semi-classical probed field $A$ agreed with \cite{JuvenSPT1}.
This confirms
our multi-kink topological term and vortex condensation mechanism do generate nontrivial SPT states.}

\section{Conclusions and discussions}
In conclusion, we have discussed the
\cblue{multi-kink topological term and vortex condensation} mechanism %and bulk dynamical effective action
for bosonic Abelian SPT states that cannot be described by \cblue{Abelian} Chern-Simons/BF actions. We have pointed out that nontrivial SPT states can be viewed as certain Higgs phases via defects proliferating in various nontrivial ways. Thus, the formalism and concepts developed in this paper \cblue{can provide further insights} for understanding the universal mechanism for bosonic SPT states, especially for those protected by non-Abelian symmetry.

Moreover, the general concept of ``hydrodynamical approach'' is applicable for fermion systems as well, if the spin-manifold is taken into account. Just like we can use \cblue{the} spin Chern-Simons theory to describe certain special Abelian fermionic SPT states\cite{SPTCS3}, the bulk effective actions beyond Chern-Simons/BF theory proposed here should also have their corresponding ``spin'' version that can describe new classes of fermion Abelian SPT states.

\cblue{ %gauge theory with {non-semi-simple} Lie algebra
The field theory based on the saddle-point approximation (detailed in Appendix \ref{non-semi-Simple}) \emph{may or may not}
fully capture the topological properties of the gapped SPT state.
%it may only describes a state close to the gapless phase transition.
However, in Sec.\ref{sec:partition-GSD}, we %are able to
show that at least for the level-1 \emph{trivial} class of our theory,
it has GSD=1 on a compact closed manifold just like the SPT state.
Moreover, so far as the SPT invariant is concerned, we confirm that the bulk SPT response theory
induced by the multi-kink topological term does reproduce the desired SPT invariant.}
%%%%
Even though our theory exhibits the so-called symmetric-self dual non-semi simple Lie algebra \cite{FigueroaO'Farrill:1995cy};
however, due to the \emph{extra set of global constraints}: Eqs.(\ref{eq:global_a}), (\ref{eq:global2}),
our theory is not equivalent to the usual gauge theory with non-semi simple Lie algebra studied in the high energy literature
(see Appendix \ref{non-semi-Simple}).
We believe our theory is \emph{unitary} and has \emph{finite ground state degeneracy on a closed manifold}.

\cblue{
Another important research direction is to study
the phase transition between superfluids and SPT states, analogous to the usual case where we have superfluid and insulator phase transition.
We will leave these further developments for future work.
}

%\cblue {\bf Juven: Can we say something about the phase transition between SPT, since in the usual case, we have superfluid and insulator transition?}

\section{Acknowledgments}
% X-G. W
This work is supported by NSF
Grant No.  DMR-1005541 and NSFC 11274192.  It is also supported by the BMO
Financial Group and the John Templeton Foundation.  Research at Perimeter
Institute is supported by the Government of Canada through Industry Canada and
by the Province of Ontario through the Ministry of Research.
We thank the IPAM workshop at University of California, Los Angeles, and its organizers for hospitality,
where the manuscript is finalized during the event.
JW is grateful to Jose Miguel Figueroa-O'Farrill, Arkady Tseytlin and Edward Witten
for their comments on the gauge theory with non-semi-simple Lie algebra.
%JW is grateful to the comments from Jose Miguel Figueroa-O'Farrill, Arkady Tseytlin and Edward Witten on
%the gauge theory with non-semi-simple Lie algebra.

\appendix

%\cblue{In the work, we will formulate the above physical picture in a more precise and more formal field theory language.}

\section{%Review and the main result}
Disorder a superfluid state into a Mott insulator or an SPT state} \label{sec:disordersf}
%Higgs phase

%\textbf{Move this subsection to later part of this paper}

To guide the readers understanding our formalism, here we briefly review this
approach using field theory (see the pioneer work\cite{{Fisher-Lee},{Dasgupta-Halperin},{Nelson}} and Ref.\cite{Zee:2003mt,Wenbook} for a field theory approach).
We plan to study SPT states for a discrete
Abelian symmetry group.  First, we will embed our discrete Abelian symmetry
group into the symmetry group of several $U(1)$ symmetries.  Instead of
starting with a discrete-symmetry breaking state, we will start with a symmetry
breaking state that break several $U(1)$ symmetries.  When we restore the
$U(1)$ symmetries, we also restore our real discrete symmetry.

The superfluid state (the $U(1)$ symmetry breaking state) in any $d$-spacetime
dimension is described by a bosonic $U(1)$ quantum phase kinetic term, whose the
partition function $\mathbf{Z}$ is:
\begin{eqnarray} \label{eq:sf_Z}
%(\partial_\mu \theta_{\text{smooth}}+\partial_\mu \theta_{\text{singular}})
\mathbf{Z}=\int [D \theta] \exp( - \int \dd^d x \, \frac{\chi}{2} (\partial_\mu \theta_{\text{s}}+\partial_\mu \theta_{\text{v}})^2 )
\end{eqnarray}
with a smooth piece $\theta_{\text{s}}$ and a singular piece $\theta_{\text{v}}$ for the bosonic phase,
and the superfluid compressibility $\chi$.
We stress that the $\theta_{\text{v}}$ is essential to capture the vortex core.
We can introduce an auxiliary field $j^\mu$ and implement the Hubbard-Stratonovich technique\cite{Zee:2003mt},
%\begin{eqnarray}
\begin{equation}
\mathbf{Z}=\int [D \theta] [D j^\mu] \exp( - \int  \dd^d x \,
\frac{1}{2\chi}(j^\mu_I)^2 - \ii  j^\mu (\partial_\mu \theta_{\text{s}}+\partial_\mu \theta_{\text{v}})).
\end{equation}
%\end{eqnarray}
By integrating out the smooth part $\int [D \theta_{\text{s}}]$, we obtain a constraint $\delta(\partial_\mu j^\mu)$ in the measure of the path integral.
We can define a generic form
$$j^\mu=\frac{1}{2 \pi (d-2)!} \epsilon^{\mu \mu_2  \dots \mu_{d} } \partial_{\mu_2} a_{\mu_3 \dots \mu_{d}},$$
with an anti-symmetric $a$ and the total spacetime dimension $d$, to satisfy this constraint.
%Equivalently,
More conveniently, in the differential form notation, the constraint is $\dd (*j)=0$ and the resolution is
$j= \frac{1}{2\pi} (*\dd a)$ with $*$ the Hodge star, with an $a$ gauge field in real values.
To disorder the superfluid, we have to make the $\theta$-angle strongly fluctuates ---
namely we should take the $\chi < \chi_c$ or $\chi \to 0$ limit \cite{Wenbook} to achieve large $(\partial_\mu \theta)^2$.
We will however drop the Maxwell term due to its irrelevancy in the renormalization group (RG) sense.
%Not the London limit of superfluid
%London limit: the scale of spatial variation is larger than the correlation length.
%
The partition function becomes:
$
\mathbf{Z}=\int [D \theta_{\text{v}}] [D a] \exp(\ii  \int
   \frac{1}{2\pi}   a  \wedge  (\dd^2 \theta_{\text{v}})  )$.
Hereafter we %may drop some irrelevant $\pm$-sign, and we
compensate the dropped $\pm$-sign by redefining the fields.
Even though naively $\dd^2=0$, due to the singularity core of $\theta_{\text{v}}$, the $(\dd^2 \theta_{\text{v}})$ can be nonzero.
Thus, $(\dd^2 \theta_{\text{v}})$ describes the vortex core density and the vortex current, which we shall denote $(\dd^2 \theta_{\text{v}})/(2\pi)=* j_{\text{vortex}}$.
In addition, the action has a symmetry of $a \to a + d \xi$, or more explicitly
$a_{\mu_3 \dots \mu_{d}} \to a_{\mu_3 \dots \mu_{d}} + \partial_{[\mu_3}  \xi_{\mu_4 \dots \mu_{d}]}$.
By Noether theorem, this symmetry leads to the conservation of the vortex current: the continuity equation
$d * j_{\text{vortex}}=0$, this implies that
$$* j_{\text{vortex}} \equiv (\dd^2 \theta_{\text{v}})/(2\pi)=\dd b/(2\pi)$$
 for some gauge field $b$.
We can thus define the singular part of bosonic phase $\dd \theta_{\text{v}}=b$ as a 1-form gauge field, to describe the vortex core,
so
\begin{equation}
\label{eq:singular_b}
\dd \theta_{\text{s}}+\dd \theta_{\text{v}}=\dd \theta_{\text{s}}+b.
\end{equation}
%{\bf \cred{The compactness of $b$ is for supporting the topological defect.}}
The partition function in the disordered state away from the superfluid, now becomes
that of an insulator state, $\mathbf{Z}=\int [D b] [D a] \exp( \frac{\ii}{2\pi} \int  b \wedge \dd a \,)$ with a topological BF action.
More explicitly, the path integral formalism shows
%\begin{eqnarray}
\begin{equation} \label{ed:Z-BF}
\mathbf{Z}=\int [D b] [D a] \exp( \ii \int
\frac{\dd^d x}{2 \pi (d-2)!} \epsilon^{\mu \mu_2  \dots \mu_{d} } b_{\mu}\partial_{\mu_2} a_{\mu_3 \dots \mu_{d}}
).
\end{equation}
%\end{eqnarray}

The Hamiltonian of Eq.(\ref{ed:Z-BF}) is zero, which describes an insulator
with an energy gap separating the ground state from excitations.  It has no
intrinsic topological order in the sense that it has a unique ground state
degeneracy (GSD, see Ref.\cite{Wen:1997ce}, this action is a level-1 BF theory
with GSD=1). This is known as the mechanism of disordering the charge while
condensing the vortices generates a trivial insulator: a Mott insulator without
SPT order.

\begin{widetext}

%\section{Derivation of the dynamical effective bulk action of type III 2+1D SPT with $Z_{N_1}\times Z_{N_2}\times Z_{N_3}$ symmetry} \label{derivation}

\section{Derivation of the dynamical effective bulk action of SPT} \label{derivation}

In the following, we list some details for deriving the internal field theory of SPT in Sec.\ref{sec:Tri-kink topological}, specifically
for type III 2+1D SPT with $Z_{N_1}\times Z_{N_2}\times Z_{N_3}$ symmetry.
We note that up to a total derivative, the tri-kink action Eq.(\ref{tri-kink})
can be simplified as:
\begin{eqnarray}
\mathcal{L}_\text{tri-kink}={1\over2}(\partial_\mu \theta^I_\s + b_\mu^I)^2+\frac{\ii }{3} C_{IJK}\varepsilon^{\mu\nu\lambda}[-3\theta^I_\s  \partial_\mu ( b_\nu^J b_\lambda^K) -3\theta^I_\s  \partial_\mu (\partial_\nu \theta^J_\s  b_\lambda^K)+ b_\mu^I b_\nu^J b_\lambda^K] + \mathcal{L}_\text{Maxwell}^b ,
\end{eqnarray}
Again, we can introduce Hubbard-Stratonovich fields $j^\mu_I$ to decouple the quadratic term as
\begin{eqnarray}
\mathcal{L}_\text{tri-kink}=\frac{1}{2}(j^\mu_I)^2-\ii \theta^I_\s  \partial_\mu j^\mu_I+ \ii b_\mu^I j^\mu_I + \frac{\ii }{3} C_{IJK}\varepsilon^{\mu\nu\lambda}[-3\theta^I_\s  \partial_\mu ( b_\nu^J b_\lambda^K)-3\theta^I_\s  \partial_\mu (\partial_\nu \theta^J_\s  b_\lambda^K) + b_\mu^I b_\nu^J b_\lambda^K] + \mathcal{L}_\text{Maxwell}^b ,
\end{eqnarray}
We further introduce Lagrangian multiplier fields \cblue{$\xi^\mu_I$ and $\lambda_\mu^I$} to decouple the $-C_{IJK}\varepsilon^{\mu\nu\lambda}\theta^I \partial_\mu (\partial_\nu \theta^J b_\lambda^K)$ term. We have:
\begin{eqnarray}
\mathcal{L}_\text{tri-kink}&=&\frac{1}{2}(j^\mu_I)^2-\ii \theta^I_\s  \partial_\mu j^\mu_I+ \ii b_\mu^I j^\mu_I + \frac{\ii }{3} C_{IJK}\varepsilon^{\mu\nu\lambda}[-3\theta^I_\s  \partial_\mu ( b_\nu^J b_\lambda^K) + b_\mu^I b_\nu^J b_\lambda^K]-\ii \theta^I_\s  \partial_\mu \xi^\mu_I \nonumber\\&&
+ \ii\lambda_\mu^I(\xi^\mu_I-C_{IJK}\varepsilon^{\mu\nu\lambda}\partial_\nu \theta^J_\s  b_\lambda^K)+\mathcal{L}_\text{Maxwell}^b
\nonumber\\&=&\frac{1}{2}(j^\mu_I)^2-\ii\theta^I_\s  \partial_\mu (j^\mu_I+\xi^\mu_I+C_{IJK}\varepsilon^{\mu\nu\lambda} b_\nu^J b_\lambda^K) + \ii b_\mu^I j^\mu_I + \frac{\ii }{3} C_{IJK}\varepsilon^{\mu\nu\lambda} b_\mu^I b_\nu^J b_\lambda^K \nonumber\\ &&+ \ii\lambda_\mu^I(\xi^\mu_I-C_{IJK}\varepsilon^{\mu\nu\lambda}\partial_\nu \theta^J_\s  b_\lambda^K)+\mathcal{L}_\text{Maxwell}^b \nonumber\\
&=&\frac{1}{2}(j^\mu_I)^2-\ii \theta^I_\s \partial_\mu \left [j^\mu_I+\xi^\mu_I+C_{IJK}\varepsilon^{\mu\nu\lambda} (b_\nu^J b_\lambda^K-\lambda_\nu^J b_\lambda^K)\right]+ \ii b_\mu^I j^\mu_I + \frac{\ii }{3} C_{IJK}\varepsilon^{\mu\nu\lambda} b_\mu^I b_\nu^J b_\lambda^K + \ii\lambda_\mu^I \xi^\mu_I +\mathcal{L}_\text{Maxwell}^b,\nonumber\\
\end{eqnarray}

Integrating out the $\theta^I_\s$ fields result in a constraint: $\partial_\mu \left [j^\mu_I+\xi^\mu_I+C_{IJK}\varepsilon^{\mu\nu\lambda} (b_\nu^J b_\lambda^K-\lambda_\nu^J b_\lambda^K)\right]=0$. From this constraint, we can write the conserved $j^\mu_I={1\over2\pi}\varepsilon^{\mu\nu\lambda}\partial_\nu a_\lambda^I- \xi^\mu_I- C_{IJK}\varepsilon^{\mu\nu\lambda} (b_\nu^J b_\lambda^K-\lambda_\nu^J b_\lambda^K)$.
Finally, we obtain:
\begin{eqnarray}
\mathcal{L}_\text{tri-kink}&=&{\ii \over2\pi}\varepsilon^{\mu\nu\lambda} b_\mu^I \partial_\nu a_\lambda^I - \frac{2\ii}{3} C_{IJK}\varepsilon^{\mu\nu\lambda} b_\mu^I b_\nu^J b_\lambda^K + \ii C_{IJK}\varepsilon^{\mu\nu\lambda} b_\mu^I \lambda_\nu^J b_\lambda^K+\frac{1}{2}\left[{1\over2\pi}\varepsilon^{\mu\nu\lambda}\partial_\nu a_\lambda^I- C_{IJK}\varepsilon^{\mu\nu\lambda} (b_\nu^J b_\lambda^K-\lambda_\nu^J b_\lambda^K)\right]^2
\nonumber\\ &&+
\frac{1}{2}(\xi^\mu_I)^2+\left[\ii (\lambda_\mu^I-b_\mu^I)-{1\over2\pi}\varepsilon^{\mu\nu\lambda}\partial_\nu a_\lambda^I+C_{IJK}\varepsilon^{\mu\nu\lambda} (b_\nu^J b_\lambda^K-\lambda_\nu^J b_\lambda^K)\right]\xi^\mu_I+\mathcal{L}_\text{Maxwell}^b ,
\end{eqnarray}

Integrating out the $\xi^\mu_I$ fields, we end up with:
\begin{eqnarray}
\mathcal{L}_\text{tri-kink}&=&{\ii \over2\pi}\varepsilon^{\mu\nu\lambda} b_\mu^I \partial_\nu a_\lambda^I - \frac{2 \ii}{3} C_{IJK}\varepsilon^{\mu\nu\lambda} b_\mu^I b_\nu^J b_\lambda^K+ \ii C_{IJK}\varepsilon^{\mu\nu\lambda} b_\mu^I \lambda_\nu^J b_\lambda^K+\frac{1}{2}\left[{1\over2\pi}\varepsilon^{\mu\nu\lambda}\partial_\nu a_\lambda^I- C_{IJK}\varepsilon^{\mu\nu\lambda} (b_\nu^J b_\lambda^K-\lambda_\nu^J b_\lambda^K)\right]^2
\nonumber\\ &&-\frac{1}{2}\left[\ii (\lambda_\mu^I-b_\mu^I)-{1\over2\pi}\varepsilon^{\mu\nu\lambda}\partial_\nu a_\lambda^I+C_{IJK}\varepsilon^{\mu\nu\lambda} (b_\nu^J b_\lambda^K-\lambda_\nu^J b_\lambda^K)\right]^2+\mathcal{L}_\text{Maxwell}^b \nonumber\\
&=& {\ii \over2\pi}\varepsilon^{\mu\nu\lambda} b_\mu^I \partial_\nu a_\lambda^I - \frac{2 \ii}{3} C_{IJK}\varepsilon^{\mu\nu\lambda} b_\mu^I b_\nu^J b_\lambda^K+\ii C_{IJK}\varepsilon^{\mu\nu\lambda} b_\mu^I \lambda_\nu^J b_\lambda^K
+\frac{1}{2}(\lambda_\mu^I-b_\mu^I)^2\nonumber\\&&+\ii(\lambda_\mu^I-b_\mu^I)\left[{1\over2\pi}\varepsilon^{\mu\nu\lambda}\partial_\nu a_\lambda^I-C_{IJK}\varepsilon^{\mu\nu\lambda} (b_\nu^J b_\lambda^K-\lambda_\nu^J b_\lambda^K)\right]+\mathcal{L}_\text{Maxwell}^b \nonumber\\
&=& {\ii \over2\pi}\varepsilon^{\mu\nu\lambda}\lambda_\mu^I \partial_\nu a_\lambda^I + \frac{\ii }{3} C_{IJK}\varepsilon^{\mu\nu\lambda} b_\mu^I b_\nu^J b_\lambda^K- \ii C_{IJK}\varepsilon^{\mu\nu\lambda} b_\mu^I \lambda_\nu^J b_\lambda^K \cblue{+} \ii C_{IJK}\varepsilon^{\mu\nu\lambda} \lambda_\mu^I \lambda_\nu^J b_\lambda^K
+\frac{1}{2}(\lambda_\mu^I-b_\mu^I)^2+\mathcal{L}_\text{Maxwell}^b,\nonumber\\
&=& {\ii \over2\pi}\varepsilon^{\mu\nu\lambda}\lambda_\mu^I \partial_\nu a_\lambda^I + \frac{\ii }{3} C_{IJK}\varepsilon^{\mu\nu\lambda} \big(\lambda_\mu^I \lambda_\nu^J \lambda_\lambda^K + %\nonumber\\&& + \frac{\ii }{3} C_{IJK}\varepsilon^{\mu\nu\lambda}
 (b_\mu^I-\lambda_\mu^I) (b_\nu^J-\lambda_\nu^J) (b_\lambda^K-\lambda_\lambda^K) \big)
+\frac{1}{2}(b_\mu^I-\lambda_\mu^I)^2+\mathcal{L}_\text{Maxwell}^b,\;\;\;\;\;
\end{eqnarray}
\end{widetext}

\section{Comments on non-semi-simple Lie algebra and topological field theory} \label{non-semi-Simple}

In Section \ref{sec:saddle-approx}, we learn that
the saddle point approximation leads us to an intrinsic field theory and a bulk dynamical theory with non-semi simple Lie algebra.
\cblue{If we write the gauge connection in terms of its gauge field components and its generators:
\begin{eqnarray}
&&\tilde{a}_\mu^\alpha T^\alpha \equiv b_\mu^I  X_I + a_\mu^I H_I^*.\\
&& (\tilde{a}_\mu^1 T^1,\tilde{a}_\mu^2 T^2, \tilde{a}_\mu^3 T^3)=
(b_\mu^1  X_1,b_\mu^2  X_2,b_\mu^3 X_3),\\
&& (\tilde{a}_\mu^4 T^4,\tilde{a}_\mu^5 T^5, \tilde{a}_\mu^6 T^6)=
(a_\mu^1  H_1^*,a_\mu^2  H_2^*, a_\mu^3 H_3^*).
\end{eqnarray}
Here $\alpha=1,\dots,6$ and $I=1,\dots,3$.}

\cblue{
The corresponding generators $H^I$ and $X_I$
satisfy:
\begin{align} \label{eq:Lie_algebra}
[H_I^*,H_J^*]=[H_I^*,X_J]=0; \quad [X_I,X_J]=C_{IJK}H_K^*,
\end{align}
where $C_{IJK}$ serves as the structure constant now.
The full Lie algebra consists of an Abelian Lie algebra $\mathcal{X}(X)$ with a central %Abelian
extension by another Abelian Lie algebra $\mathcal{H}^*(H^*)$.
Here $\mathcal{X}(X)$ contains the set of generators $X_I$, and $\mathcal{H}^*(H^*)$ contains the set of generators $H_I^*$. }

For the specific case of level-1 Chern-Simons theory in Sec.\ref{sec:partition-GSD}, we are able to show the GSD=1.
However, for the general level-$k$ case, the structure of the phase space volume is changed (see for example, Appendix \ref{GSD}).
This Appendix is meant to provide some word of caution to prevent us from making a stronger claim that the
Chern-Simons theory with this non-semi simple Lie algebra is exactly the dynamical Dijkgraaf-Witten field theory we look for,
unless we carefully specify the global constraints analogous to Sec.\ref{sec:partition-GSD}.

The particular type of the non-semi simple Lie algebra we derived in Eq.(\ref{eq:Lie_algebra}) is in the class of
\emph{symmetric-self dual Lie algebra} \cite{FigueroaO'Farrill:1995cy}.
Even if the Killing form $\kappa_{ab}$ degenerates, we can replace the $\kappa_{ab}$ by an invariant nondegenerate symmetric bilinear form
$\mathcal{K}^{G}_{a \alpha'}$ if it satisfies the criteria below.

For a Lie algebra given by $[T_a,T_b]= f_{ab}{}^c T_c$, the structure constant $f_{ab}$ satisfies
the Jacobi identity:
$f_{bc}{}^{d} f_{ad}{}^{e} +f_{ca}{}^{d} f_{bd}{}^{e}+f_{ab}{}^{d} f_{cd}{}^{e}=0$.
The Killing form as a bilinear matrix in the adjoin representation can be determined from the structure constant.
\begin{eqnarray}
\kappa_{ab}=\kappa(T_a,T_b)=-\Tr(T_a, T_b) %=-\Tr(f_{a \alpha}{}^\beta\, f_{b \beta}{}^\alpha)
=-\sum_{\alpha, \beta} f_{a \alpha}{}^\beta\, f_{b \beta}{}^\alpha.
\end{eqnarray}
The Killing form is called degenerate, if
there exists a nonzero generator $T'$ such that
$\kappa(T',T)=0$ for any $T$.

In the Euclidean spacetime, we have a Chern-Simons theory:
\begin{eqnarray}
\label{eq:L-general}
L={\frac{\ii }{4\pi} } \epsilon^{\mu\nu\rho} \mathcal{K}^{G}_{a \alpha'} \Big(  \mathcal{A}^a_{\mu}(x) \partial_\nu \mathcal{A}^{\alpha'}_{\rho}(x) \nonumber\\
+\frac{1}{3} f_{bc}{}^{a} \mathcal{A}^{\alpha'}_{\mu}(x) \mathcal{A}^b_{\nu}(x) \mathcal{A}^c_{\rho}(x)\Big).
\end{eqnarray}
Even if the Killing form is degenerate,
as long as this $({\mathcal{K}_{}^{G}})_{IJ}$ can be found, the $({\mathcal{K}_{}^{G}})_{IJ}$ can replace the degenerate Killing form
to make sense of the Chern-Simons theory Eq.(\ref{eq:L-general}) with
the symmetric-self dual Lie algebra.

The $({\mathcal{K}_{}^{G}})_{IJ}$ is a symmetric non-degenerate invariant bilinear form, %$\Omega_{IJ} \equiv$
constrained by:
\begin{eqnarray}
{f_{a \ell}}^i ({\mathcal{K}_{}^{G}})_{bi}+ {f_{ab}}^i  ({\mathcal{K}_{}^{G}})_{  \ell i} =0
%{f_{a \ell}}^i \Omega_{bi}+ {f_{ab}}^i  \Omega_{  \ell i} =0
\end{eqnarray}

The finite and infinitesimal gauge transformations are:
\begin{eqnarray}
&&\mathcal{A}_\mu \to \mathcal{A}_\mu^U=  U^{-1} (\mathcal{A}_\mu +  \partial_\mu ) U= e^{- \alpha^a T_a} (\mathcal{A}_\mu + \partial_\mu ) e^{ \alpha^a T_a}, \nonumber\\
&&\mathcal{A}^a_\mu(x) \to  (\mathcal{A}^a_\mu(x) +   f_{bc}{}^{a}   \mathcal{A}^b_\mu(x) \alpha^c(x) +  \partial_\mu \alpha^a(x)). \;\;\;\;\;\;\,
\end{eqnarray}

The Lie algebra we find out in Section \ref{sec:saddle-approx} is a sub-algebra of the most generic symmetric-self dual Lie algebra \cite{FigueroaO'Farrill:1995cy}:
\begin{eqnarray}
&&[X_a,X_b]= i {f^{(X)}_{ab}}{}^c X+ i {f^{(H^*)}_{ab}}{}^\alpha H^*_\alpha,\\ %\\ \;
&&[H_a,H_b]=i {f^{(H)}_{ab}}{}^c H_c,\\
&&[X_a,H_b]=i {f^{(xH)}_{ab}}{}^c X_c,\\
&&[H_a,H_b^*]=-i {f^{(H)}_{ac}}{}^b H_c^*,\\
&&[X_a,H^*_\alpha]= [H^*_\alpha,H^*_\beta]=0.
\end{eqnarray}
Notice that the subalgebra spanned by $\mathcal{X}(X)$ and $\mathcal{H^*}(H^*)$, is the Abelian extension of
$\mathcal{X}(X)$ by $\mathcal{H^*}(H^*)$.
The full algebra is the semidirect product of $\mathcal{H}(H)$ by this Abelian extension.
The particular non-semi-simple symmetric-self dual Lie algebra in Eq.(\ref{eq:Lie_algebra}) is nilpotent, non-abelian, non-reductive and solvable.
The corresponding Lie group is non-compact.

Our theory in Appendix \ref{sec:partition-GSD} is a special case such that the GSD is still 1 which can describe the gapped SPT.
Due to the non-compact Lie group, however, it is likely the generic gauge theory of symmetric-self-dual Lie algebra
can capture an infinite degenerate gapless phase instead of a phase with finite topological degenerate ground states.
The concern of (non-)unitarity has been investigated, for example, in Ref.\onlinecite{Tseytlin:1995yw}.

We believe that the generic difference between our SPT path integral and the usual non-semi-simple-Lie-algebra gauge theory
is the set of global constraints: Eqs.(\ref{eq:global_b}), (\ref{eq:global_a}), (\ref{eq:global2}).
For our SPT path integral, the global constraints lead to the finite ground state degeneracy,
for the usual non-semi-simple-Lie-algebra gauge theory, the ground state degeneracy can be infinite.
It is possible a more generic theory can describe a state close to the potential gapless phase transition between superfluids, symmetry-breaking states
and SPT/topologically ordered states.
We will leave the further investigation open for future work.

%for

\section{Counting the degenerate zero modes} \label{GSD}
% by using the path integral

\subsection{GSD for a gapped system with a 0+1D topological term} \label{sec:GSD1}

We first review a simple ground state degeneracy (GSD) calculation by counting the zero mode for a 0+1D system.
Namely we will count the volume of the phase space volume,
\begin{eqnarray}
\GSD = \text{the volume of the phase space},
\end{eqnarray}
up to some normalization factor.

The first system we consider is described by a Berry phase term ${{\mathcal{L}^0}}= \dot{X} P$. %in the partition function:
On one hand, in the path integral formalism, we have a partition function:
\begin{eqnarray}
\mathbf{Z}=\int [DX] [DP] \exp[  {\ii} \; {k} \int  \dot{X} P],
\end{eqnarray}
$ \dot{X}=\partial_0 X$ is the time derivative $X$.

On the other hand, in the quantum operator formalism, we have the commutator $[X, \frac{\partial{\mathcal{L}^0}}{\partial  \dot{X}}]=\ii$:
\begin{eqnarray}
[X,P]=\ii \frac{1}{k}.
\end{eqnarray}
$X$ and $P$ are some matrix operators acting on the 0+1D space.  Here we will consider
a compact phase space, so that the phase space volume is finite. In particular, without losing generality, the identification we assume is $X \sim X+ 2\pi$ and $P \sim P+1$. Since the Hamiltonian is essentially $H=\dot{X} P -{\mathcal{L}^0}=0$,
the system seems to be trivial without kinetic terms or potential terms.
However, there can be degenerated ground states. All ground states $\Psi$ satisfy $H \Psi =0$.
But these $\Psi$ may not be all independent. To count the GSD thus to count the independent degree of freedom,
we can construct a generic ground state $\Psi$
in terms of the function of $X$ if we choose $X$ as the basis:
\begin{eqnarray}
\Psi(X)= \sum_{n \in \mathbb{Z}} c_n e^{ \ii  n X}
\end{eqnarray}
The form is obtained by satisfying the constraint: $\Psi(X)=\Psi(X+2 \pi)$ as $X \sim X+2 \pi$.
The $2 \pi$ shift in the exponent will not affect the form of the $\Psi(X)$ function.
On the other hand, by doing the Fourier transformation, we can transform the $X$ basis to the $P$ basis via
$\tilde{\Psi}(P)=\int  e^{ \ii {k} P X} \Psi(X) \; dP$. Up to some normalization factor, this yields,
\begin{eqnarray}
\tilde{\Psi}(P)= \sum_{n \in \mathbb{Z}} c_n \delta(kP+ n)
\end{eqnarray}
Meanwhile, the form satisfies the constraint: $\tilde{\Psi}(P)=\tilde{\Psi}(P+1)$ as $P \sim P+1$.
This implies that $c_n \delta(kP+  n)=c_{n-{k}} \delta(kP+k+  ({n-{k}}))$.
This means that
\begin{eqnarray}
c_n =c_{n-{k}}
\end{eqnarray}
with ${k} \in \mathbb{Z}$.
The volume of the phase space is $|k|$.
We have $|k|$ independent degenerate ground states determined by $k$ independent coefficients,
thus $\GSD=|k|$.
The strategy for this example is basically the same as the approach in Ref.\onlinecite{Wen:1997ce}.

\subsection{Compactification and Quantization}

For the later convenience, we now set up a relation between the constraint of {compactification and quantization} using an angular rotational system
as an example, with
the angle $\Theta$ and the angular momentum $L$. First, $\Theta$ is compactified and identified via:
$$
\Theta \sim \Theta + 2\pi.
$$
The compactness of $\Theta$ leads to the quantization or the discretization of its dual variable $L$, in order to have $e^{\ii \Theta L}$ stays invariant as
$\Theta \to \Theta + 2\pi$. That means, the quantization is
$$
\Delta L = 1.
$$
On the other hand, if we consider the angle $\Theta$ is also discretized as rotor angle with
$$
\Delta \Theta = \frac{2 \pi}{N},
$$
then this quantization must come from the compactification of $L$, with
$$
L \sim L + N.
$$
In short, due to the constraint of {compactification and quantization}, we have a set of relations:
\begin{eqnarray}
\Theta \sim \Theta + 2\pi \Leftrightarrow \Delta L = 1,   \label{logic1} \\
L \sim L + N \Leftrightarrow \Delta \Theta = \frac{2 \pi}{N}. \label{logic2}
\end{eqnarray}
The volume of the phase space is $N$.
It can be counted in $\Theta$-space as well as in $L$-space as
$(2\pi/\Delta \Theta)=(N/\Delta L)=N$.

\subsection{GSD for a gapped system at the 0+1D edge of 1+1D SPTs} \label{sec:GSD2}

After the previous simple first part of calculation, in the second part, we consider the 0+1D edge of 1+1D SPT.
The system we consider is described by a Berry phase term in the partition function for the path integral formalism:
\begin{eqnarray}
\mathbf{Z}=\int [D \varphi^1] [D \varphi^2] \exp[ \frac{\ii}{2} \int  C_{IJ} \varphi^I \partial_0 \varphi^J],
\end{eqnarray}
with $C_{12}=\frac{p_{\text{II}}N_1N_2}{2\pi  N_{12}}$.
%Quantum mechanically,

On the other hand, for the canonical quantization with quantum opearators, the commutation relation satisfies
\begin{eqnarray}
[\varphi^1, \varphi^2]=\frac{\ii}{C_{12}}=\frac{2\pi \ii N_{12}}{p_{\text{II}}N_1N_2}.
\end{eqnarray}
To well-define the denominator for the trivial class ${p_{\text{II}}}=0$, the trivial class's ${p_{\text{II}}}$ is identified as ${p_{\text{II}}}=N_{12}$.
We may define the conjugate variables as $[\varphi^1, P_{\varphi^1}]=[\varphi^1, {C_{12}}\varphi^2]=\ii$
and $[\varphi^2, P_{\varphi^2}]=[\varphi^2, -{C_{12}}\varphi^1]=\ii$.

{\bf The 1st approach:} The compactified size of $\varphi^1$ and $\varphi^2$ is no larger than $2\pi$,
\begin{eqnarray} \label{compact1}
\varphi^1 \sim \varphi^1 + 2\pi, \;\;\; \varphi^2 \sim \varphi^2 + 2\pi.
\end{eqnarray}
The quantization and the discreteness of these rotor clock is %at least as
no smaller than:
\begin{eqnarray} \label{quantize1}
\Delta \varphi^1 = \frac{2\pi}{N_1}, \;\;\; \Delta \varphi^2 = \frac{2\pi}{N_2}.
\end{eqnarray}
Due to the conjugation relation, following the logic of Eq.(\ref{logic1}), the compactness in Eq.(\ref{compact1}) of $\varphi^1 \sim \varphi^1 + 2\pi$ leads to
$\Delta P_{\varphi^1}={C_{12}} \Delta \varphi^2 =1$. Similarly, the compactness of $\varphi^2$ leads to
$\Delta P_{\varphi^2} ={C_{12}} \Delta \varphi^1=1$.
Namely, the quantization can be:
\begin{eqnarray} \label{quantize_b}
\Delta \varphi^1 = \frac{2\pi  N_{12}}{p_{\text{II}}N_1N_2}, \;\;\; \Delta \varphi^2 = \frac{2\pi N_{12}}{p_{\text{II}}N_1N_2}.
\end{eqnarray}
On the other hand, following the logic of Eq.(\ref{logic2}), the quantization Eq.(\ref{quantize1}) implies the possible compactness size of
$P_{\varphi^1}$ and $P_{\varphi^2}$ as: $P_{\varphi^1} \sim P_{\varphi^1}+N_1$ and $P_{\varphi^2} \sim P_{\varphi^2}+N_2$. namely,
\begin{eqnarray} \label{compact_b}
\varphi^1 \sim \varphi^1 + \frac{2\pi  N_{12}}{p_{\text{II}}N_1}, \;\;\; \varphi^2 \sim \varphi^2 +  \frac{2\pi  N_{12}}{p_{\text{II}}N_2}.
\end{eqnarray}

To construct the refined phase space, we need to take the largest quantization size in the discretized lattice among
Eq.(\ref{quantize1}) and Eq.(\ref{quantize_b}), and the smallest compactification size among
Eq.(\ref{compact1}) and Eq.(\ref{compact_b}).
This means that we will require Eq.(\ref{quantize1}) and Eq.(\ref{compact_b}):
\begin{eqnarray}
&&\Delta \varphi^1 = \frac{2\pi}{N_1}, \;\;\; \Delta \varphi^2 = \frac{2\pi}{N_2}, \nonumber\\
&& \varphi^1 \sim \varphi^1 + \frac{2\pi  N_{12}}{p_{\text{II}}N_1}, \;\;\; \varphi^2 \sim \varphi^2 +  \frac{2\pi  N_{12}}{p_{\text{II}}N_2}. \nonumber
\end{eqnarray}
Therefore the phase-space-volume counting from both $\varphi^1$-space and its dual space, $\varphi^2$-space,
is both $\frac{2\pi  N_{12}}{p_{\text{II}}N_1} /\Delta \varphi^1=\frac{2\pi  N_{12}}{p_{\text{II}}N_2} /\Delta \varphi^2=\frac{N_{12}}{p_{\text{II}}}$.
However, $\frac{N_{12}}{p_{\text{II}}}$ may not be integer in general. We will need to multiply a minimal factor on the size of the phase space until it becomes
an integer. This means that in general we will multiply it by the minimal phase factor $\frac{p_{\text{II}}}{\gcd(p_{\text{II}},{N_{12}})}$ until we have an integer size of phase volume:
%\begin{eqnarray}
$\frac{N_{12}}{\gcd(p_{\text{II}},{N_{12}})} =\frac{N_{12}}{p_{\text{II}}} \cdot \frac{p_{\text{II}}}{\gcd(p_{\text{II}},{N_{12}})}$.
%\end{eqnarray}
The phase-space-volume counting from both $\varphi^1$-space and $\varphi^2$-space results in
\begin{eqnarray}
\GSD=\frac{N_{12}}{\gcd(p_{\text{II}},{N_{12}})}.
\end{eqnarray}

It is straightforward to construct the functional $\Psi(\varphi^1)$ and its Fourier transformation
 $\tilde{\Psi}(\varphi^2)$ with a number of $\frac{N_{12}}{\gcd(p_{\text{II}},{N_{12}})}$ independent coefficients as in Sec.\ref{sec:GSD1}.\\

{\bf The 2nd approach:} We can verify this GSD result from an alternative viewpoint, by considering the projective representation of the symmetry
group $G=Z_{N_1} \times Z_{N_2}$:

We propose the symmetry generators as
\begin{eqnarray}
&&{S_{\varphi^1}=e^{\ii N_1\varphi^1 \frac{p_{\text{II}}}{N_{12}}}},\\
&&{S_{\varphi^2}=e^{\ii N_2\varphi^2  \frac{p_{\text{II}}}{N_{12}}}},
\end{eqnarray}
in order to have the symmetry generators invariant under the shift over a full compactification size
$\varphi^1 \to \varphi^1+\frac{2\pi  N_{12}}{p_{\text{II}}N_1}$ and $\varphi^2 \to \varphi^2+\frac{2\pi  N_{12}}{p_{\text{II}}N_2}$.
Namely, our choice is guaranteed to satisfy: $S_{\varphi^1}({\varphi^1})=S_{\varphi^1}({\varphi^1}+\frac{2\pi  N_{12}}{p_{\text{II}}N_1})$
and $S_{\varphi^2}({\varphi^2})=S_{\varphi^2}({\varphi^2}+\frac{2\pi  N_{12}}{p_{\text{II}}N_2})$.
Our choice also obeys the $Z_{N_1}$ and $Z_{N_2}$ symmetry: $(S_{\varphi^1}){}^{N_1}=(S_{\varphi^2}){}^{N_2}=1$ when we have
impose the discretization as Eq.(\ref{quantize1}).

One can check that
$
S_{\varphi^I} (\int \dd t \mathcal{L}_\text{edge}^0)  S_{\varphi^I}^{-1}=(\int \dd t \mathcal{L}_\text{edge}^0) + 2 \pi \cdot \text{integer},
$
so the partition function $\mathbf{Z}=\int D {\varphi^1} D {\varphi^2} e^{-\int \dd t \mathcal{L}_\text{edge}^0} \ $is invariant under the symmetry transformation $S_{\varphi^I}$.
To calculate the ground state degeneracy at the 0+1D edge, we can study the projective representation of the symmetry group acting on the zero energy modes, we find:
{
\begin{eqnarray}
S_{\varphi^1} S_{\varphi^2}=e^{-\frac{2\pi \ii p_\text{II}}{ N_{12}}} S_{\varphi^2} S_{\varphi^1}.
\end{eqnarray}
}
If $p_\text{II}=0$, the symmetry generators are commutative, so it can be written as a linear representation; and the GSD=1.
In general, the symmetry generators are not commutative, so it shall be written as a higher dimensional matrix representation.
If $\gcd(p_\text{II},N_{12})=1$, it is the $Z_{N_{12}}$ Heisenberg algebra and it requires a $N_{12}$ dimensional representation.
This implies the 0+1D edge mode of the ground state has $\GSD={N_{12}}$, %a ${N_{12}}$-fold degeneracy,
consistent with the edge mode physics analysis via the dimensional reduction approach in Ref.\cite{Wang:2014tia}.
If $\gcd(p_\text{II},N_{12}) \neq 1$, we can reduce the rank of the representation matrix to a smaller rank,
we rewrite
$$
S_{\varphi^1} S_{\varphi^2}=e^{-{2\pi \ii  }  \frac{1}{ \frac{N_{12}}{\gcd(p_{\text{II}},{N_{12}})}} \frac{p_\text{II}}{\gcd(p_{\text{II}},{N_{12}})} } S_{\varphi^2} S_{\varphi^1}.
$$
In this way, we obtain a relative-prime factor $\frac{p_\text{II}}{\gcd(p_{\text{II}},{N_{12}})}$, and the GSD is the reduced rank of the matrix representation of the symmetry generators:
\begin{eqnarray}
\GSD=\frac{N_{12}}{\gcd(p_{\text{II}},{N_{12}})}. \nonumber
\end{eqnarray}

%\subsection{GSD for a gapped system with the 0+1D kink on 1+1D domain-wall edge of 2+1D SPTs}

We have shown the degenerate zero modes happening on the 0D edge of 1D bulk SPT.
In general, if we create various symmetry breaking domain wall to gap the gapless boundary mods of the higher dimensional boundaries,
we can study the zero modes trapped at the gapped domain wall via the dimensional reduction approach. As an example,
we can look into the  0+1D kink on a 1+1D domain-wall edge Eq.(\ref{edge})
of 2+1D bulk SPTs.
%$\mathcal{L}_\text{edge}^1=\frac{\ii }{3} \frac{1}{(2 \pi) 2!} \frac{n_1 N_2 N_3\; p_{\text{III} }}{N_{123}}\varepsilon^{\mu\nu}  \partial_\mu \varphi^J \partial_\nu \varphi^K$,
This result is consistent with Refs.\onlinecite{JuvenSPT1} and \onlinecite{Wang:2014tia}.
\color{black}

\end{document}